# Dynamically Stable Radiation Pressure Propulsion of Flexible Lightsails for Interstellar Exploration

Ramon Gao[1], Michael D. Kelzenberg[1], and Harry A. Atwater*

California Institute of Technology, Pasadena CA 91125

[1]These authors contributed equally to this work.

*Corresponding author: haa@caltech.edu

**Abstract**

Lightsail spacecraft, propelled to relativistic velocities via photon pressure using high power density laser radiation, offer a potentially new route to space exploration within and beyond the solar system, extending to interstellar distances. Such missions will require meter-scale lightsails of submicron thickness, posing substantial challenges for materials science and engineering. We analyze the structural and photonic design of flexible lightsails, developing a mesh-based multiphysics simulator based on linear elastic theory, treating the lightsail as a flexible membrane rather than a rigid body. We find that flexible lightsail membranes can be spin stabilized to prevent shape collapse during acceleration, and that certain lightsail shapes and designs offer beam-riding stability despite the deformations caused by photon pressure and thermal expansion. Excitingly, nanophotonic lightsails based on planar silicon nitride membranes patterned with suitably designed optical metagratings exhibit both mechanically and dynamically stable propulsion along the pump laser axis. These advances suggest that laser-driven acceleration of membrane-like lightsails to the relativistic speeds needed to access interstellar distances is conceptually feasible, and that fabrication of such lightsails may be within the reach of modern microfabrication technology.

## Introduction

The concept of harvesting radiation pressure to propel spacecraft dates to at least some 400 years ago, when Kepler observed that the gas tails of comets point away from the sun as if blown by a solar wind (1). The physics of radiation pressure became known when Maxwell published his theory of electromagnetism in the 19th century, giving rise to formal development of the concept of solar lightsails by Tsiolkovsky, Tsander, and others in the early 20th century (2). Efforts to field solar lightsail spacecraft have led to recent successes including the JAXA IKAROS (3), NASA NanoSail-D (4), and the Planetary Society LightSail missions (5).

Whereas sunlight provides a relatively weak force for accelerating spacecraft in Earth's vicinity (~10 μN/m$^2$ for a perfect reflector at 1 AU), far greater accelerating forces can be produced if a high power density laser is focused onto a lightsail. Simple analysis suggests that laser-propelled lightsails can in principle be accelerated to relativistic velocities, offering a promising pathway for interstellar exploration using ultralight space probes (6–8). Due in part to the announcement of the Breakthrough Starshot Initiative in 2016, which seeks to enable this capability within the next generation (9, 10), recent investigations have explored the viability of laser-driven lightsails as a basis for interstellar spacecraft propulsion (8, 11–13). A major challenge for such lightsails is the need to maximize reflectance while minimizing weight and limiting optical absorption to extremely low values, prompting multilayer or nanophotonic designs (14–18). Given the extreme rates of acceleration and the distances over which this acceleration will occur, it is necessary that such lightsails are designed to be structurally and dynamically stable, such that they can be propelled along the pump laser beam optical axis (19–33) with shape-stable configuration. Several designs for rigid- or constrained-body beam-riding lightsails have been proposed, but to date no studies have considered the mechanical and beam-riding stability of meter-scale unsupported flexible membranes for interstellar propulsion. Notably, to achieve the target velocity of ~0.2*c*, the Starshot mission concept calls for ~1 g lightsail that is several meters in diameter; thus the membrane must be on the order of 100 atomic layers thick on average, including all framing or stiffening, suggesting that the flexibility of the lightsail must be taken into account in its design.

Here we consider the selection of materials, the structural and photonic design, and dynamic mechanical stability of flexible lightsail membranes, to investigate whether interstellar lightsail spacecraft can be realized with real materials, considering their finite stiffness and strength. We identify key material properties required for relativistic flexible lightsails, then develop a multiphysics simulation approach to explore the deformation and passively stabilized acceleration of spinning flexible lightsails with either specular scattering concave shapes or flat membranes with embedded metagrating nanophotonic elements.

**Table 1.** Figures of merit for mechanical strength of candidate lightsail materials.

| Material | Young's Modulus E (GPa) | Tensile strength $\sigma_T$ (GPa) | Density $\rho$ (g/cm$^3$) | Thickness *t* for 0.1 g/m$^2$ (nm) | $D_{max}$ burst for 67 Pa (m) 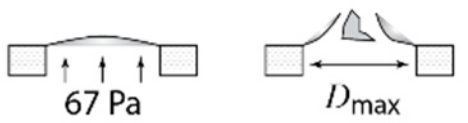  | $f_{max}$ spin for 10 m$^2$ (Hz) 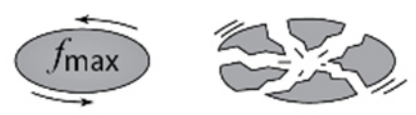  |
|---|---|---|---|---|---|---|
| Silicon (111 surf.) | 169 | 2.1 | 2.33 | 43 | 1.3 | 130 |
| Diamond (PECVD nano) | 750 | Up to 7.5 | 2.52 | 31 | 1.7 | 180 |
| $SiO_2$ (tempered glass) | 66 | Up to 1.0 | 2.42 | 41 | 0.66 | 90 |
| $Si3N_4$ (LPCVD film) | 250 | 6.4 | 2.7 | 37 | 4.6 | 210 |
| $MoS_2$ (bilayer) | 200-240 | 21 | 5.02 | 20 | 19 | 300 |
| | | | | | | |
| Aluminum | 72 | 0.50 | 2.80 | 36 | 0.18 | 60 |
| Polyimide | 2.5 | 0.069 | 1.42 | 70 | 0.10 | 30 |

## Materials considerations

The Breakthrough Starshot Initiative (9) has challenged a global community of scientists and engineers to design a ~1-gram interstellar probe that will travel 4.2 light-years to reach Proxima Centauri B, the nearest known habitable-zone exoplanet, within ~20 years of launch, as well as the necessary propulsion, communication, and instrumentation systems for such a mission. To accelerate the spacecraft to the required speed of ~0.2*c*, a ~10 m$^2$ lightsail weighing ~1 g would be propelled by an earth-based laser at incident power densities approaching ~10 GW/m$^2$, experiencing ~10,000 Gs of acceleration for ~1000 seconds (8, 10). A lightsail suitable for this mission must address immense engineering obstacles that will challenge the limits of materials science and engineering. One such challenge is that the lightsail must have reasonably high optical reflectance to produce thrust from the accelerating beam, yet must exhibit near-zero optical absorption (~1 ppm or less) and high thermal emissivity to prevent overheating. Recent studies have identified a handful of dielectric and semiconductor materials as potentially viable candidates (11–13), and nanophotonic designs made from these materials (or their combinations), have been reported to offer favorable reflectance, low absorption, and high emissivity (14–17, 34). In addition to achieving suitable optical properties over a wide temperature range, lightsail materials and designs must offer adequate mechanical strength and stiffness to endure the acceleration conditions necessary for interstellar propulsion. Table 1 shows key room-temperature mechanical properties and structural performance metrics for several of the candidate lightsail materials identified by previous studies, as well as for two common solar sail materials: aluminum and polyimide. More detailed properties are references provided in Table S1.

Among the candidates are several bulk crystalline dielectrics and semiconductors, including Si, quartz ($SiO_2$), and diamond, which are hard, brittle, and have among the highest moduli and theoretical strengths of known bulk materials. Despite this such materials are rarely used in bulk structural applications, and are notorious for brittle failure in tension due to cracks initiated at surface defects. In practice, attainable specimen strength is limited almost entirely by the ability to fabricate device structures with defect-free surfaces. Although many decades of materials science and engineering have enabled each of these materials to achieve remarkable degrees of purity and scale of manufacture, present-day technology has yet to produce pure, defect-free, submicron-thick membranes over 10 m$^2$ areas.

Two-dimensional crystals represent another class of candidate materials, among which $MoS_2$ appears particularly promising for lightsail applications owing to its high strength and refractive index (15). The reported tensile strength for micron-scale suspended membranes of mono- or bi-layer $MoS_2$ is nearly three times higher than that of any other material listed in Table S1. (35) Understanding the achievable strength and optical transparency of $MoS_2$ films fabricated over large or non-planar surfaces, at relevant layer thicknesses, and at elevated temperatures, is of considerable interest.

Another interesting class of materials for lightsail development is that of amorphous or nanocrystalline deposited thin films, including silicon nitride. Such thin-film materials are widely used for modern MEMS. Promisingly, sub-micron thickness silicon nitride membranes have been fabricated at wafer scale, and further patterned with photonic crystal designs for near-unity reflectance (36, 37). Ultralow extinction coefficients on the order of $10^{-6}$ at near-infrared wavelengths can be achieved with high-stress stoichiometric silicon nitride ($Si_3N_4$), which is commonly employed in MEMS and cavity optomechanics applications (38, 39). With favorable mechanical properties including high modulus and tensile strength, and potential research synergies between the fields of cavity optomechanics and optical levitation, $Si_3N_4$ is a particularly promising candidate material for lightsail development.

Ultimately, considerable effort will be required to develop any suitable materials system(s) to the scale of manufacture required for the interstellar lightsails proposed by the Starshot initiative, and careful consideration must be paid to the resulting mechanical and optical properties of the lightsail materials over a wide range of operating temperatures.

**Stability considerations**

In addition to possessing adequate optical and mechanical properties to endure the forces and optical intensities of the propulsion laser beam, the overall lightsail design must provide for adequate stability during acceleration. Our work addresses two key aspects of stability: *beam-riding stability*, the ability of the lightsail to follow along the beam axis without external guidance, and *structural stability*, the ability of the lightsail to survive the acceleration sequence without collapse, disruptive deformation of its shape, or tensile failure of its constitutive materials. These challenges and potential solutions are depicted schematically in Figure 1.

It is tempting to assume that the lightsail should be propelled by a beam of uniform laser intensity, to minimize thermal gradients and force nonuniformities that could distort the lightsail shape. This is the operating regime for solar sails, which navigate via active local control of solar reflectance or other attitude control mechanisms (1, 2). However, uniform plane-wave illumination is impractical for laser-propelled interstellar lightsails, as it would require a laser source of inconceivable power and aperture area to overcome diffraction of the beam over the extreme distance of acceleration. Assuming the propulsion laser would be constructed no larger than necessary to achieve the target mission velocity, the system must operate at or near the diffraction limit during the final phase of acceleration. We therefore restrict our study of beam-riding stability to static, weakly-focused low-order Gaussian beam intensity profiles. Other beam profiles such as higher-order Gaussian beams or doughnut beams (21, 25) may be useful at earlier stages of acceleration when the propulsion system is not limited by diffraction.

Passive beam-riding stability is necessary for relativistic lightsail acceleration, because it is not feasible to provide closed-loop propulsive corrections by modulating or adjusting the propulsion beam in response to observations of the lightsail, owing to the large acceleration distances and final lightsail velocity. Limited by the speed of light, the round-trip delay between lightsail observation and the arrival of corrective modulation from the laser source in an active feedback loop would range up to several minutes at the end of the acceleration phase, whereas non-beam-riding lightsails can veer off course on a timescale of milliseconds. Additionally, atmospheric turbulence and practical technological limitations will cause at least some perturbation to the desired position and profile of the beam (40). Thus, although some initial prescriptive corrective actions may be feasible from the laser source, the lightsail itself must ultimately be capable of aligning its acceleration trajectory to the beam axis without ground-based intervention, based solely on the local beam gradient. The challenge of steering the spacecraft then becomes primarily that of correctly pointing and slewing the direction of the ground-based laser source during acceleration.

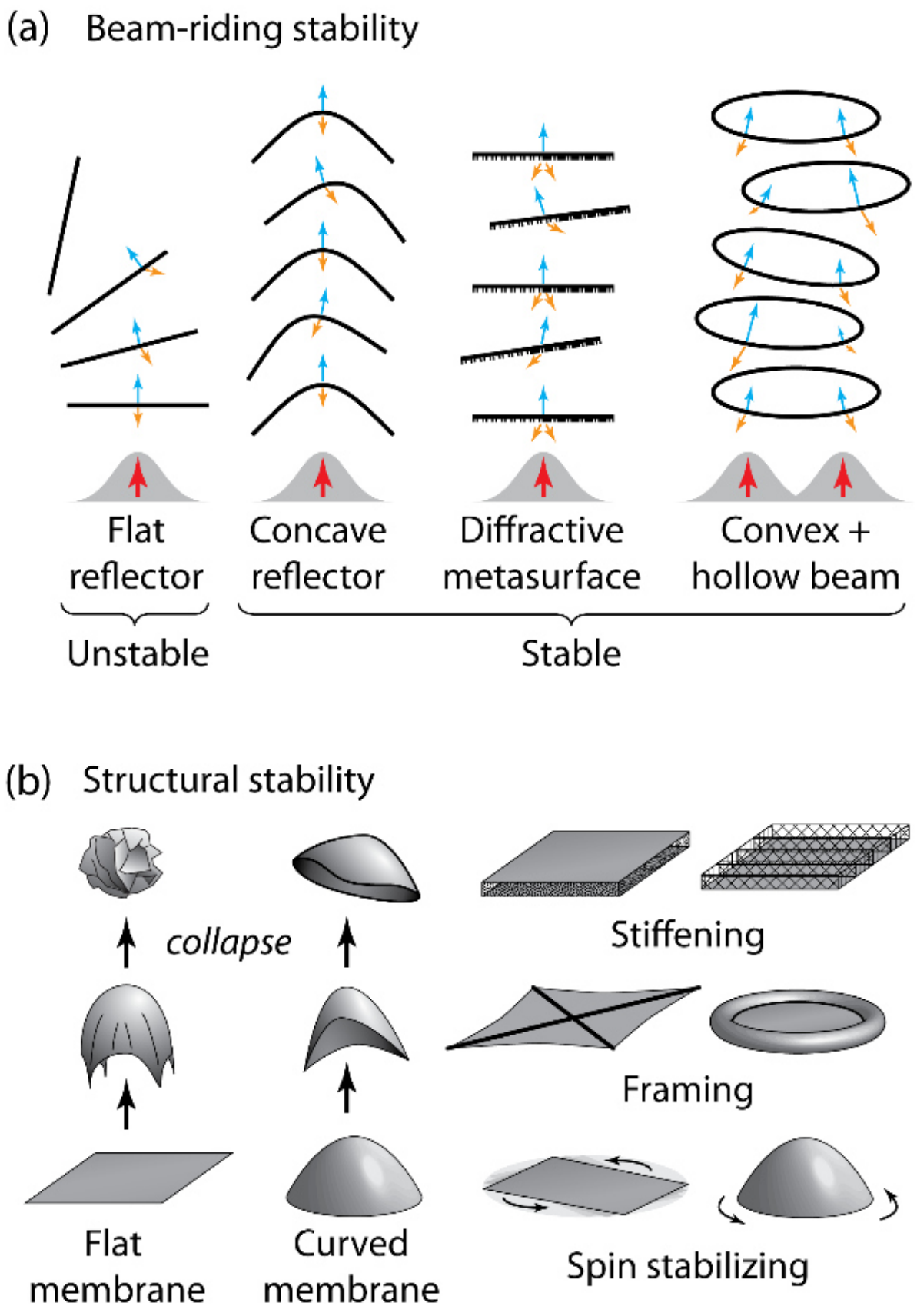

**Figure 1.** Conceptual illustrations of design approaches. Designs for achieving *(a)* beam-riding stability, and *(b)* structural stability, in lightsail membranes. In panel *(a)*, the red arrow depicts the accelerating beam position, the orange arrows indicate the direction of reflected light, and the blue arrows indicate the force of radiation pressure.

A simple lightsail structure such as a flat specularly reflective disk is dynamically unstable and will eventually tilt and veer away from the beam. Several approaches to achieving beam-riding stability are depicted in Fig. 1A. Certain geometrically concave reflector shapes, including cones (21–23), hyperboloids (19), paraboloids, and other parametric shapes (32) have been predicted to offer stable beam-riding behavior, while other normally unstable convex shapes such as spheres can follow a stable trajectory by using more complex higher-order beam profiles (21). In addition to shaped specular lightsails, non-specular surfaces can be employed to produce restoring forces and torques, even for flat lightsails, by tailoring asymmetric optical properties to effect transverse forces (18, 24–28, 33). Non-specular surfaces have been developed for solar lightsails to achieve greater maneuverability including enhanced lateral and rotational forces (41).

Our present study addresses only marginal (undamped) beam-riding stability, in which the lightsail exhibits bounded, oscillatory displacement and tilting about the beam axis in response to a finite beam-lightsail misalignment, during acceleration. Continuous perturbations to the beam-lightsail alignment during propulsion, e.g., due to atmospheric turbulence, can cause the oscillatory motion of the lightsail spacecraft to grow in magnitude, which could eventually cause marginally stable lightsails to escape the beam. Furthermore, for nonrigid structures, and flexible membranes in particular (42), the energy buildup in acoustic modes (shape distortions) could also destabilize or overstress the lightsail. Therefore, interstellar lightsails will likely require either active or passive means of damping their beam-riding oscillations and shape vibrations to achieve asymptotically stable propulsion along the desired cruise trajectory. Passive damping approaches might include the use of structures with damped internal degrees of freedom (31), employing nonlinear optical materials (30), or utilizing materials with highly varying temperature-dependent optical properties to enable hysteresis of the restoring forces. Active optical control surfaces for improved beam-riding stability have been demonstrated for solar lightsails (43), but developing such control surfaces to operate under the extreme beam intensities and low mass budget proposed for interstellar lightsail propulsion remains an unsolved challenge.

Turning our attention to structural stability, the interstellar lightsail must be capable of surviving the acceleration forces without collapsing upon itself or experiencing mechanical failure. This is a substantial challenge for the Starshot concept, which calls for meter-scale lightsail membranes of average thickness below 100 nm. Table 1 shows the allowable average thickness for each membrane type. This is not intended to suggest that lightsails should be constructed from uniformly thick continuous membranes, or to impose an upper limit for structural thickness. Optimized lightsail designs will likely incorporate multiple materials (16) and complex spatial patterning, e.g., perforations (15, 17, 30, 37) or optical resonators (18, 24, 25, 33, 34), so as to maximize reflectance, emissivity, and tensile strength. However, with such limited mass budget, the finite strength and structural rigidity of the lightsail must be considered.

A prior study addressed tensile strength requirements by treating the lightsail as a rigid parametrically shaped shell, finding that certain surface curvature ranges minimize stress (44). Another recent study presented 2D analytic and finite-element models of deformation instabilities in uniformly illuminated lightsail membranes (45). In the absence of external constraints, the behavior of unsupported or loosely supported flexible membranes subject to nonuniform forces is considerably complex (42).

In general, thin unsupported membranes will collapse and crumple upon themselves when subject to focused laser propulsion, as depicted in Fig. 1B. A curved surface offers greater structural rigidity than a flat membrane, while also conferring the benefits of improved stress distribution that make thin curved shells useful in structural applications. However, open concave shapes such as cones and paraboloids are still prone to collapsing by elongation, an intrinsic instability for such shapes. Structural reinforcement such as framing could be added, but only at the cost of reducing the membrane mass. Potential approaches for structural reinforcement include microlattices (46), gas-filled envelopes (47, 48), annular tensioning, fractal supports (49), tensegrity structures (50), or lamination with low-density or corrugated backing layer(s). Ultimately, given mass and material constraints, even a structurally rigidified lightsail will likely deform during acceleration, potentially changing the distribution of stress within the membrane or altering its beam-riding properties. An additional challenge for any structural materials is that the proposed lightsail membranes are generally partially transparent. Thus, even if placed behind the lightsail surface, the frame or backing materials may still be exposed to a high laser intensity, limiting materials selection.

As an alternative to structural support, spin-stabilization may be employed to prevent shape collapse. This effectively rigidifies the lightsail via inertial tensioning, and also gyroscopically stabilizes the lightsail to resist tilting, all while avoiding the added mass and complexity of structural reinforcement. For this reason, our work to date has focused on spin stabilized lightsails. However, spin-stabilization greatly complicates the dynamics of the lightsail, particularly for flexible membranes which are prone to complex instabilities (42), and is not necessarily effective for all structures under all conditions. Perhaps most counterintuitively, gyroscopic effects can disrupt the beam-riding behavior of certain lightsail designs that would be dynamically stable under non-spinning (rigid-body) conditions, particularly in the case of angular misalignment between the beam axis and the spin axis (21, 22). Thus, the use of spin-stabilization to prevent shape collapse in ultrathin flexible lightsails can be a challenging design objective.

To provide first-order insights into the general viability of constructing large-area structurally stable lightsails from the candidate materials, we have defined two figures of merit in Table 1. The first is the *stationary burst diameter* ($D_{max}$), which is the maximal diameter at which a flat circular membrane of areal density 0.1 g/m$^2$, rigidly clamped at its perimeter, can sustain a pressure of 67 Pa applied to one side without rupturing (51). This is the effective photon pressure of 10 GW/m$^2$ illumination, assuming unity reflectance. Practical lightsail designs may have lower reflectance; may incorporate multiple materials or inhomogeneous patterning, owing to the need to optimize tradeoffs between reflectance, thermal properties, and strength; and would need to operate at substantially elevated temperatures. $D_{max}$ is thus only intended to serve as an order-of-magnitude indicator of the viability of large-area perimeter-supported membranes for this application. The assumed 'stationary' perimeter constraint provides an overestimate of the required membrane tensile strength, since any viable perimeter structure would not be stationary, but instead must have an extremely small mass that would accelerate along with the lightsail. But interestingly, even this simplified calculation suggests that while conventional solar lightsail materials such as aluminum and polyimide (Kapton) are far too weak to span meter-scale areas between structural supports, some candidate membrane materials ($Si_3N_4$ and $MoS_2$) are in principle strong enough to span 10 m$^2$ areas ($D_{max}$ > 3.6 m) with perimeter support only – even in the stationary case. This is an encouraging conclusion for the development of structurally supported lightsails.

The second figure of merit in Table 1 is the *maximum spin speed* ($f_{max}$) at which a flat, 10 m$^2$ circular membrane could be spun without rupturing due to tensile failure. This is relevant because, for the designs considered here, relatively high spin speeds are required to produce both shape stability and beam-riding stability, often approaching the materials' tensile limits. The viability of spin stabilization depends on the spin speed, the acceleration conditions, and the specific design of the lightsail. For this reason, we have developed multiphysics numerical simulation methods to investigate the dynamic stability of flexible lightsails.

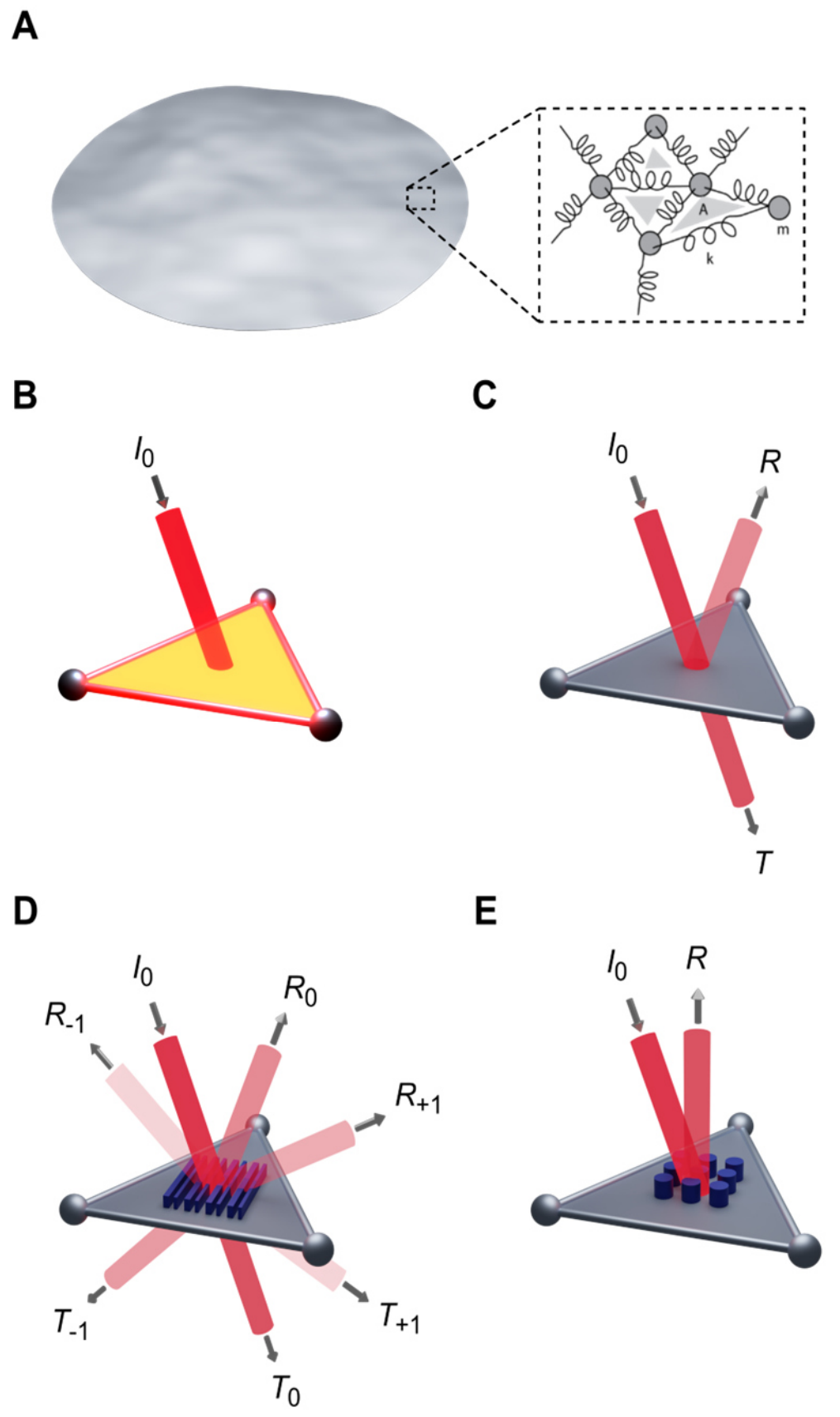

**Figure 2.** Modeling flexible lightsails and light-matter interaction with a mesh-based time-domain simulator. *(A)* Ultrathin and meter-scale lightsails and their deformations can be modeled by a mesh comprising masses *m* (nodes) connected by springs with stiffnesses *k* (edges), enclosing triangles of area *A*. Light-matter interactions are calculated for each mesh triangle based on discretization of the incident light as localized beam $I_0$. Modeled behaviors include *(B)* absorption of light and thermal emission, which heat and cool the structure, driving heat flow, thermal expansion, and changes in material properties; *(C)* specular reflection and transmission of light, producing photon pressure, and in some cases, causing reflected light to impinge other triangles; *(D)* optical diffraction from periodic wavelength-scale surface patterning, producing transverse directional forces from photon pressure, and *(E)* optical wavefront shaping such as beam steering with subwavelength optical metasurfaces.

## Mesh-based simulator for flexible lightsails

To model realistic flexible lightsail membranes of various shapes and optical designs, a triangular surface mesh is constructed (Fig. 2A). Each vertex is assigned a mass based on the local membrane thickness, the area of the adjoining triangles, and the material density. Elastic behavior of the membrane is captured by the edges, each of which is assigned a linear elastic coefficient (i.e., spring constant) based on the mesh geometry and local material properties. This approach omits the negligible bending stiffness and the specific shear modulus of the material, but provides reasonable first-order insights into the behavior of ultrathin membranes under tensile loading, which is the predominant type of loading in lightsail applications. In future efforts, non-isotropic material properties and the full elastic behavior of the lightsail material(s) could be considered.

Light-matter interactions are evaluated over each enclosed triangular mesh element, where incident light produces photon pressure forces, optical absorption heats the lightsail, and thermal radiation cools the lightsail (Fig. 2B). The heating, cooling, and optical forces calculated at each triangular element are distributed to the adjoining nodes, which represent the temperature distribution, momentum, and shape of the structure. Thermal conduction is calculated along the mesh edges based on the local material properties and mesh geometry, whereas temperature is calculated at each node based on its mass and the specific heat of the material. As the temperature distribution is known throughout the structure, we also include the effects of linear thermal expansion, which contributes to thermal strain.

In the simplest type of optical interaction, the force of photon pressure acting on a triangular element is governed by the effect of specular reflection from the surface (Fig. 2C), with the resulting force occurring normal to the surface. The photon pressure is calculated based on the local beam intensity, the relative polarization, the incidence angle to the surface, and the local membrane properties. Future efforts could also consider the optical effects of local temperature, strain, or the time-varying state of active control surfaces, as well as beam profiles that vary in time or distance from the source. Our present work has studied only the first seconds (up to 10 s) of acceleration following an initial beam-lightsail misalignment, which is adequate for observing marginally stable behavior over many periods of oscillation, determining steady-state temperature distributions, and identifying many types of instabilities. The present model does not address relativistic effects necessary to model the full acceleration duration to interstellar mission velocities.

In the next section, we first assume constant values for the reflectance, transmittance, and correspondingly absorptance to model the basic behavior of curved and flat specular lightsails. Then, we will introduce improvements to the optical calculations, including angle-dependent reflectance and absorption based on Fresnel coefficients, and considering the effects of multiple reflections of light within concave curved lightsail. Finally, we present simulations of non-specularly reflecting surfaces such as diffractive metagratings (Fig. 2D), which allow flat lightsails to achieve beam-riding stability. With future work, this basic simulation approach could be adapted to study lightsails made from optical metasurfaces (Fig. 2E) with a wide range of optical behaviors.

To simulate acceleration of the lightsail, and to assess its apparent stability, we implement a finite-difference time-domain approach wherein we calculate the forces and heat flow acting at each mesh vertex, then evaluate the resulting changes in position, velocity, and temperature over a time step $\Delta t$. With sufficiently small $\Delta t$, we can simulate the propagation of membrane vibrational modes, and can obtain reasonable predictions of the lightsail dynamic behavior during the initial acceleration phase. Thermal and mechanical membrane failures can be detected when a nodal temperature exceeds a threshold value or when the strain in an edge exceeds the tensile limit of the material.

**Dynamics of flexible curved lightsails**

Fig. 3 depicts the simulated behavior of flat versus curved (paraboloid) lightsails and the effects of spin stabilization, using optical and mechanical properties roughly corresponding to a 43 nm thick Si membrane (0.1 g/m$^2$) whose properties are parameterized at room temperature. We first consider a 1-meter diameter flat lightsail, illuminated by a $\lambda = 1550$ nm Gaussian beam profile, with 4 GW/m$^2$ peak intensity and a 0.5-meter beam waist, offset by 80 mm from the center of the lightsail. This lightsail size was chosen as a compromise between computational cost and the desire to simulate large macroscopic structures with reasonable mesh accuracy. Because these lightsails are smaller than 10 m$^2$ in area as proposed for Starshot, they can be spun faster than the values shown in Table 1; the unloaded maximum spin speed for the flat membrane in this case is ~470 Hz. Without spin stabilization, the flat lightsail membrane is structurally unstable and collapses upon itself as expected. Spin stabilization ($f_{\text{spin}} = 135$ Hz) prevents the lightsail from collapsing, but lacking any means for beam-riding stability, the lightsail quickly veers away from the beam axis. A paraboloid shape can offer beam-riding stability according to rigid-body calculations, but in the flexible mesh simulation, the membrane quickly becomes elongated and collapses upon itself. With inadequate spin stabilization ($f_{\text{spin}} = 90$ Hz), the shape collapse is delayed but not prevented, in this case leading to tensile failure. With adequate spin stabilization ($f_{\text{spin}} = 135$ Hz), the shape remains stable, and beam-riding stability is achieved throughout the 1 s duration of the simulation. Animations of all five cases are available in Supplementary Video 1.

To facilitate comparison, all lightsails in Fig. 3 have the same surface area and thus the same total mass. As a result, the paraboloid lightsails are smaller in diameter, and thus accelerate more slowly than the flat lightsails due to their smaller aperture area. Therefore, a drawback of deeply curved shapes is that they tend to be heavier than flat lightsails of the same aperture area and thickness. Also, the sloped peripheral surfaces of the paraboloids do not propel the lightsail along the *z* direction as efficiently, since some of the photon pressure is directed radially. Light reflected from these edge areas might in fact impinge somewhere on the opposite side of the lightsail, thus imparting additional photon pressure there, potentially affecting the acceleration and stability of the lightsail. We thus improved our simulation code by considering multiple reflections within the lightsail using a simplified raytracing approach, and by calculating reflectance and absorption based on Fresnel coefficients, thus better modelling the angle dependence of light interaction (Fig. 4). Animations of these and other raytracing-based simulations are shown in Supplementary Video 2.

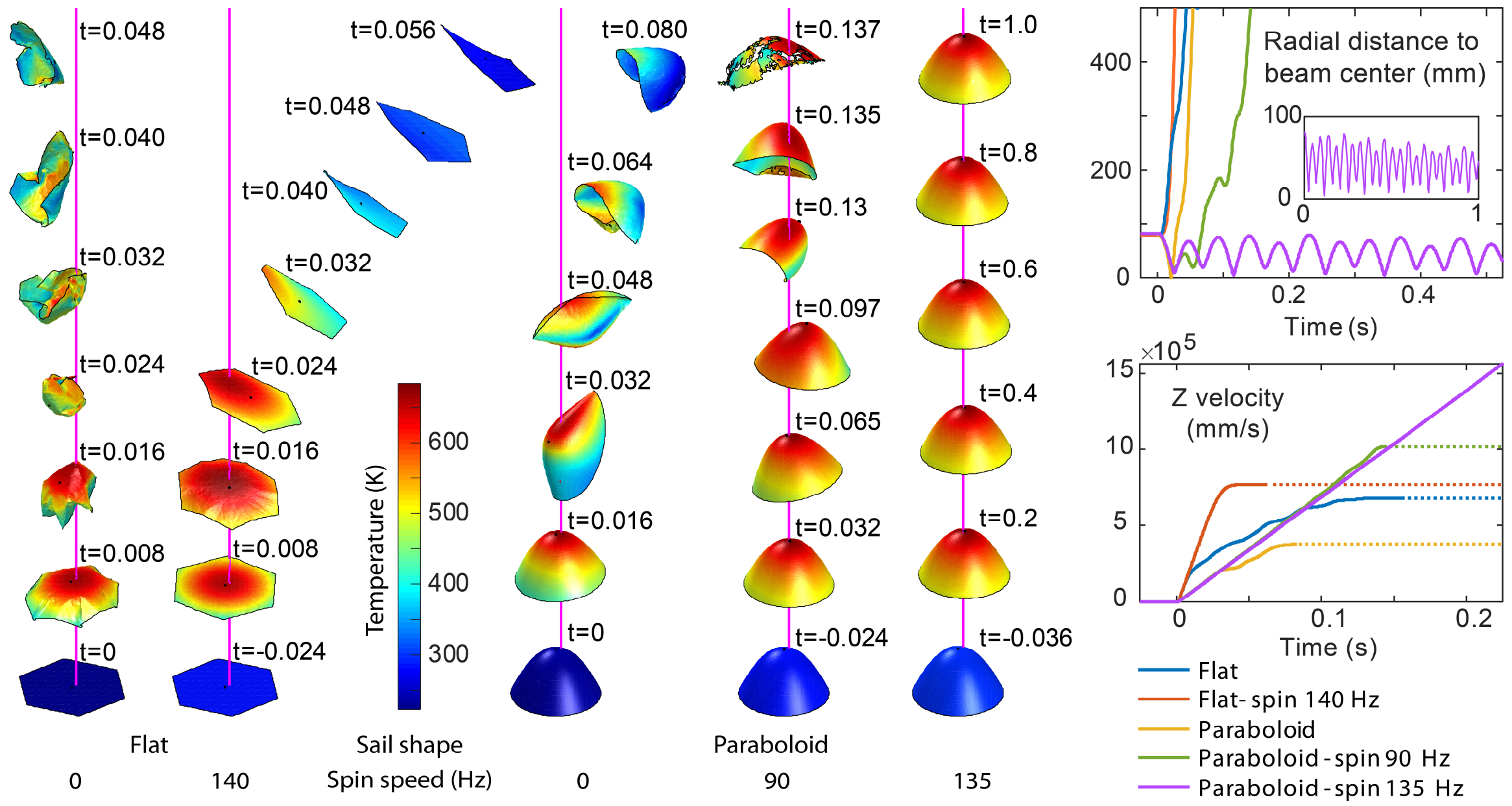

**Figure 3.** Simulation results for flat versus curved specular lightsails, with and without spin stabilization. Illumination is in the $+z$ direction starting shortly after $t = 0$, with a Gaussian profile ($I_0 = 4$ GW/m$^2$, $R_{waist} = 0.5$ m, $\lambda = 1.55$ µm), offset by 80 mm from the initial lightsail centers. *Left:* Surface renderings show temperature, shape, and lateral position of each lightsail at the indicated times during simulation. Surface shading was applied to enhance depiction of shape. The vertical magenta lines show the beam centerlines. All lightsail images appear at the same scale; however, their vertical positions have been shifted for presentation. *Right plots*: The distance between the lightsail center of mass and the beam centerline (above), and the lightsail $z$ velocity (below), plotted versus time. Animations of all five simulations are available as Supplementary Video 1.

Fig. 4A compares the shape and temperature behavior of a 1-meter diameter spin-stabilized paraboloid lightsail representing a 43-nm thick Si membrane, with and without the effects of multiple reflections within the lightsail, with simulation conditions being otherwise the same as for the stabilized paraboloid shown in Fig. 3. Due to the modest reflectivity of silicon (0.45 for $\lambda = 1550$ nm at normal incidence), the effects of reflected light can substantially disrupt the lightsail stability. Considering only the effects of the incident light beam, the lightsail trajectory appears stable, similar to that shown in Fig. 3; but upon introducing the effects of secondary reflections, the lightsail shape and trajectory become unstable. While the secondary reflections do increase the total photon pressure on the lightsail, resulting in faster acceleration, reflected light striking the opposite side of the lightsail counteracts the restoring forces and torques produced by the first reflection, thus destabilizing the lightsail. Also evident from the temperature profiles is the localized heating caused by the focusing of reflected light, with the peak temperature increasing from ~700 K to ~1000 K.

Increased temperatures are problematic for lightsails because materials generally weaken or decompose at elevated temperatures. An upper temperature limit may be imposed by material sublimation or decomposition, as even small amounts of material loss could substantially weaken or alter such thin lightsails (15). If we limit the mass loss to 1%, for a 1 g, 10 m$^2$ lightsail for 1000 s, literature values predict a limiting temperature of ~1300 K for crystalline Si (52), suggesting that the projected temperatures above are acceptable. However, for semiconductor materials, free-carrier absorption increases dramatically with temperature as the bandgap narrows, which may lead to a thermal runaway situation at a much lower threshold temperature. Furthermore, two-photon absorption may trigger thermal runaway above certain laser intensities, regardless of initial temperature. A recent analysis of an optimized Si-based nanophotonic lightsail estimated the threshold temperature for thermal runaway to be only 400–500 K, and placed an upper limit on beam intensity at ~5 GW/m$^2$ (53). Thus, our simulations predict unsurvivable temperatures for Si membranes, and unsurvivable light intensities in the regions of focused secondary reflections.

Nonetheless, we can conclude that spin-stabilization can prevent shape collapse of flexible curved lightsails. While multiple reflections within deeply curved lightsails can increase the acceleration rate, they also increase the risk of localized hotspots from focused light and can also reduce or disrupt beam-riding stability. However, this only affects curved shapes which are deep enough to encounter multiple reflections over the range of tilt angles and shape deformations experienced during acceleration. Another challenge is that curved lightsail shapes would likely be more difficult to fabricate at the meter scale, and for crystalline materials, would introduce weaknesses at joints, grain boundaries, or wherever weaker crystal planes are exposed. We are thus motivated to investigate flat membranes as an alternative to curved shapes, owing to likely easier fabrication and scale-up, and to the lack of internal secondary reflections within the lightsail. However, we note that not all curved shapes are destabilized by internal reflections, and that shallower spin-stabilized curved shapes can achieve stability without encountering conditions that produce secondary internal reflections (21) (Supplementary Video 2).

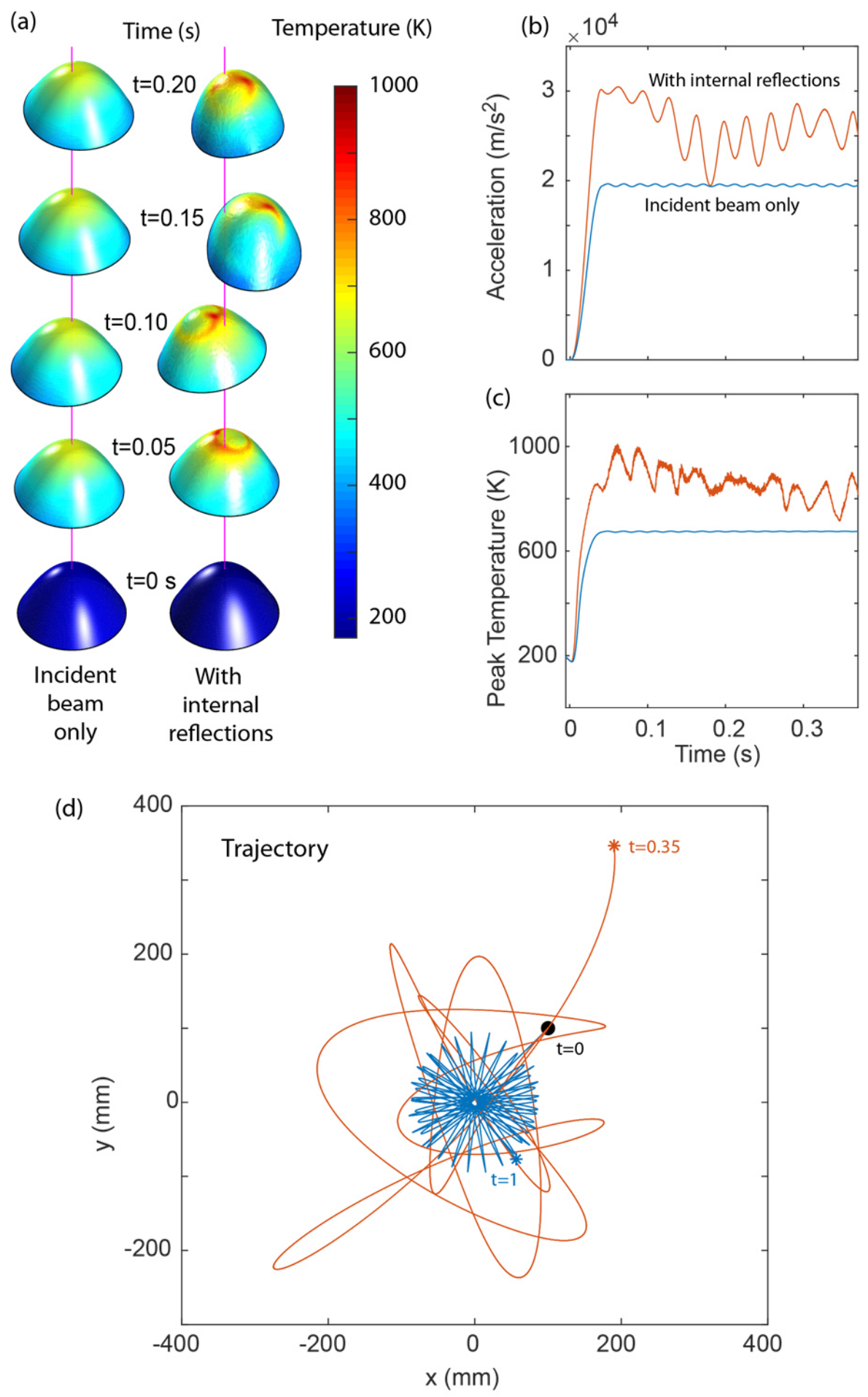

**Figure 4.** Effects of multiple internal light reflections within spin-stabilized flexible paraboloid lightsail. Simulated shape *(a)*, acceleration *(b)*, peak temperature *(c)*, and trajectory *(d)* of a 1-m diameter paraboloid lightsail, with and without the effects of internal light reflection within the lightsail. Paraboloid lightsails and acceleration conditions are similar to those in Fig. 3. Animations of these and other raytracing-based simulations are available as Supplementary Video 2.

Since Si exhibits thermal runaway at relatively low threshold temperatures and beam intensities (53), we turned our attention to a different material for the flat lightsails. Even if radiative cooling of Si-based lightsails could be improved, using any material with such a low runaway threshold temperature appears problematic, since any local defect, contamination, or brief localized focusing of light exceeding the two-photon absorption threshold, could initiate catastrophic thermal runaway spreading across the entire lightsail. $Si_3N_4$ is used extensively in other high-temperature applications, and its larger optical bandgap (~5 eV) and lower free-carrier absorption are attractive. Furthermore, amorphous $Si_3N_4$ films of excellent optical quality can be deposited using LPCVD, suggesting an easier route for fabrication over large or complex surfaces (36, 37). A drawback to $Si_3N_4$ is its relatively low refractive index ($n \sim 2$), resulting in lower reflectance and less efficient diffraction.

It is difficult to estimate the practical limiting temperature for $Si_3N_4$ lightsails based on its properties reported in literature, owing to the diversity of its applications, the varying stoichiometry, density, and stress produced by chemical vapor deposition methods, and the relative complexity of the N-Si system at high temperatures. As an upper limit, we estimate the temperature at which vacuum decomposition would occur (again choosing a threshold of 1% decomposition over 1000 s) to be ~1600 K, based on decomposition rates for crystalline powders of $Si_3N_4$ (54). Practical thermal limits would likely be much lower, as the decomposition evolves nitrogen, leaving elemental silicon at the material surface, which could dramatically increase optical absorption and lead to thermal runaway. Other high-temperature risks include weakening, changes to stress distribution, activation of traps or defects, or crystallization of the material. Further experimental measurements are necessary to accurately determine the limiting temperatures and power densities for $Si_3N_4$ lightsails.

**Optical design for passive stabilization of flat lightsails**

Passive stabilization of lightsail dynamics requires the presence of restoring forces and torques. The previously discussed concave curved shapes achieve this via their shape alone, but flat specular lightsails cannot achieve beam-riding stability because specular reflection only produces forces normal to the surface. One approach to obtain beam-riding designs for flat lightsails is to make use of engineered optical anisotropy. In diffractive gratings with symmetric unit cells, such optical anisotropy can be achieved with nematic liquid crystals (55). Alternatively, optical anisotropy can be created by designing asymmetric diffractive metagratings, e.g., with the unit cells comprising two resonators of dissimilar widths (24, 33). In such structures, anisotropic scattering of incident light into the grating diffraction orders manifests in optical forces transverse to the membrane. Moreover, optical metasurfaces comprising subwavelength scatters in the form of disks (18), blocks (25), or spheres (29) can be used to shape the wavefronts of scattered light, redirecting incident photon momentum in anomalous ways to produce beam-riding stability.

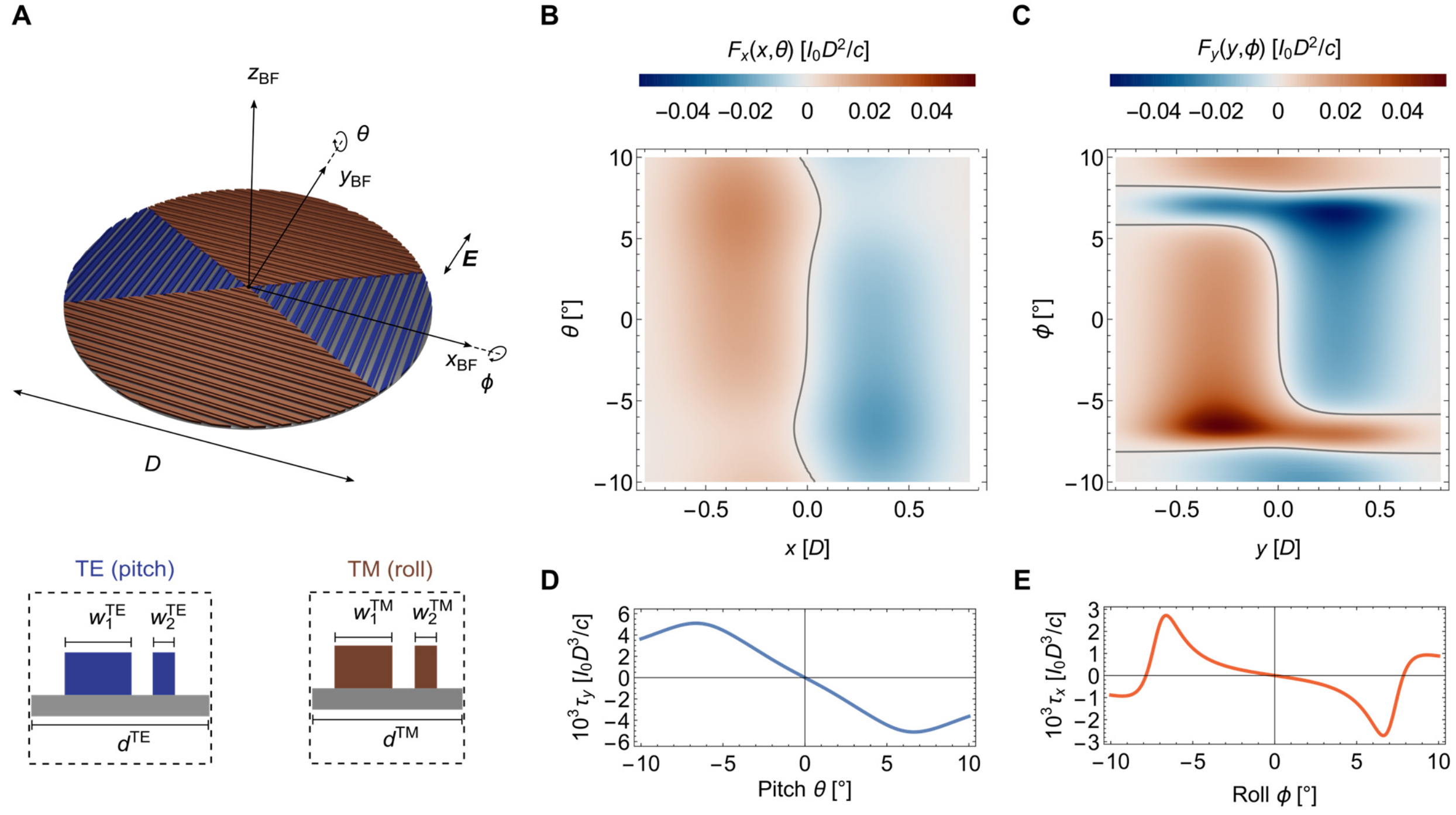

**Figure 5.** Optical design and restoring forces and torques on a metagrating-based silicon nitride lightsail. *(A)* Conceptual illustration of lightsail patterning with two mirror-symmetrically arranged metagrating designs. *(B)* Unit cells of metagrating design operating for TE polarization (top) and TM polarization (bottom). *(C)* Optical force $F_x$ as a function of normalized translation $x$ and tilt $\theta$, being self-stabilizing due to $dF_x(x, \theta = 0)/dx < 0$ between ±0.35$D$. Contour line depicts the equilibrium position, $F_x = 0$. *(D)* Optical force $F_y$ as a function of normalized translation $y$ and angle $\phi$ being self-stabilizing due to $dF_y(y, \phi = 0)/dy < 0$ between ±0.3$D$. Contour lines depict $F_y = 0$. *(E)* Optically induced torques $\tau_y(\theta)$ and $\tau_x(\phi)$, evaluated at the beam center. Self-restoring behavior ($d\tau_y(\theta)/dy < 0$, $d\tau_x(\phi)/d\phi < 0$) occurs over the approximate range of ±6.5° in both directions.

We describe stable designs for flat lightsails by designing asymmetric diffractive metagratings, patterned from $Si_3N_4$ as shown in Fig. 5. A specifically designed pair of mirror-symmetrically arranged metagratings can passively stabilize translations and rotations along one axis (24, 33). Consequently, we employ two distinct and perpendicularly arranged metagrating designs to enable stabilization of translations along both $x$ and $y$, and rotations $\theta$ about $\mathbf{y}_{BF}$ (pitch) and $\phi$ about $\mathbf{x}_{BF}$ (roll). As shown in Fig. 5A, a circular lightsail is partitioned into four sectors, forming two orthogonal pairs of symmetrically opposed wedges. We assume a linearly polarized incident beam, with its electric field aligned with the body-frame $y$-axis $\mathbf{y}_{BF}$. Thus the blue sectors (1/6 of the lightsail area) experience transverse-electric (TE) polarization, and the brown sectors (1/3 of the lightsail area) experience transverse-magnetic (TM) polarization, and the specific asymmetric metagratings for each sector (Fig. 5B) provide stabilizing forces and torques for their respective design planes and polarization. For spin-stabilized lightsails, we assume that the beam polarization rotates synchronously with the spinning lightsail. Electromagnetic simulations were performed to determine the optical response of metagrating unit cells.

For a laser propulsion wavelength of $\lambda = 1064$ nm, we identified self-stabilizing metagrating designs using linearized stability analysis. While non-spinning designs are marginally stable if the eigenvalues of the Jacobian matrix derived from the lightsail equations of motion are purely imaginary, for spinning lightsails as linear-time periodic systems, we must employ Floquet theory to assess stability of the designs (56, 57). Specifically, our chosen unit cell designs for lightsails spinning at 120 Hz produce absolute values of eigenvalues equal to 1, i.e., $|\lambda_i| = 1$, which is a sufficient and necessary condition for marginal stability. We study the initial acceleration of our marginally stable lightsail designs, subject to an initial alignment error, which allows us to verify the beam-riding stability predicted by linearized rigid-body Floquet analysis, and importantly, to investigate whether these spinning sails retain their beam-riding stability when the assumption of rigidity is removed.

The two metagrating designs each support $m = \pm 1$ diffraction orders in addition to the specular order in reflection and transmission. Asymmetry in the intensities of the diffracted orders provides the mechanism for lateral restoring forces, while asymmetry in the angular dependence of optical thrust provides the mechanism for restoring torques. Assuming a Gaussian beam with a width equal to 40% of the lightsail diameter, i.e., $w = 0.4D$, we calculated the normalized optical forces and torques induced on a rigid lightsail of the proposed design, over a range of incidence angles ($\theta$, $\phi$) and translational offsets ($x$, $y$). These induced forces do not depend on acceleration distance $z$ and yaw tilt $\psi$ because we neglect beam divergence and assume synchronous rotation of the polarization. Stabilizing behavior is evident from the negative slopes of $F_x$ and $F_y$ versus $x$ and $y$, respectively, with zero crossings (equilibrium positions, indicated by gray isolines) present near the beam center ($x, y = 0$) over the full ±10° range of plotted tilt angles $\theta$ and over a ~ ±5° range of roll angles $\phi$, respectively (Fig. 5C, 5D). The relative insensitivity of lateral equilibrium position to tilt angle appears to be beneficial for improving stability in the spinning case.

Restoring torques limit angular rotation relative to the optical axis, although the situation is less straightforward for the spinning case. Beam-center optical torques about $x$ and $y$ are shown in Fig. 5E, exhibiting stabilizing polarity and derivative over a ±6.5° range of pitch and roll. While the TE metagrating provides a larger torque about $y$, the TM metagrating yields slightly stronger optical forces along $y$. We note that $\tau_x(\phi)$ is markedly nonlinear beyond ~ ±1.5°, which restricts conclusions drawn from linear stability analysis to this angular range. Rotations beyond ±1.5° will give rise to nonlinear dynamics, resulting in possible coupling to and between distinct frequency components. Our time-domain numerical simulations allow this behavior to be studied by considering the full angle-resolved optical response of the metagratings.

**Dynamics of metagrating-based lightsail**

To verify our predictions about dynamical stability of *rigid* lightsails patterned with the composite metagrating design reported here, we numerically solved the equations of motion. The dynamics of *flexible* lightsails with the same metagrating motif were also simulated using our mesh-based modeling approach. The lightsail diameter is $D$ = 1 m, for which the chosen composite metagrating design yields a total mass of $m$ = 0.867 g. A Gaussian propulsion beam with a peak intensity of $I_0$ = 1 GW/m$^2$ and a width of $0.4D$ = 40 cm was assumed.

We present here an exemplary case of passive stabilization of a flexible metagrating lightsail, in which an initial translational offset of $x = y = 5$ cm in the lightsail position relative to the beam optical axis and an initial (pitch and roll) tilt of $\theta = \phi = -2°$ was assumed (Fig. 6). In the Supplementary Information, we also present results for passive stabilization of a flexible metagrating lightsail being only initially displaced (Fig. S3), but not tilted relative to the beam optical axis. Snapshots of the flexible lightsail position, orientation and shape every 0.5 s are shown in Fig. 6A; an animation of the simulation is available as Supplementary Video 3. For the studied duration of $t = 5$ s, the lightsail oscillates about the beam axis while remaining relatively flat and level, with no visibly apparent shape distortion thanks to the sufficiently large tensioning forces arising from spin-stabilization. Due to the finite absorptivity of $Si_3N_4$, the center region of the lightsail reaches a maximum temperature of 959 K. In contrast, the peripheral area remains significantly cooler (Fig. S4a), heating up to a maximum temperature 489 K. The slower heat up process on the edge of the lightsail can be attributed to limited heat transport from the hot center of the lightsail, owing to the low thermal conductance of the silicon nitride membrane. Thermal conduction dominates over direct absorption as a source of heating in the peripheral areas of the lightsail, due to the underfilling laser beam. The peak temperature appears sufficiently below the vacuum decomposition temperature of $Si_3N_4$ (54), although this temperature is likely too hot for most payloads. Increasing the assumed hemispherical emissivity of 0.1 for thin $Si_3N_4$ membranes would be desirable, for example with additional metasurface designs for selective thermal radiation in the mid-infrared regime, or addition of other material layers (15, 16, 58). Our simulation predicts a maximum strain of 0.091% in the $Si_3N_4$ membrane (Fig. S5A). With a Young's modulus of 270 GPa, such strain translates to a tensile stress of approximately 246 MPa, which is > 40 times lower than the reported 6.4 GPa tensile limit of $Si_3N_4$. (Table 1). Therefore, a meter-sized flexible lightsail is expected to exhibit mechanical stability in its propulsion phase despite being subject to large thermal gradients, spin tensioning, and nonuniform beam intensity.

Examining the trajectory of flexible and rigid lightsails indicates that their motion is bounded and thus the dynamics appear to be marginally stable as expected

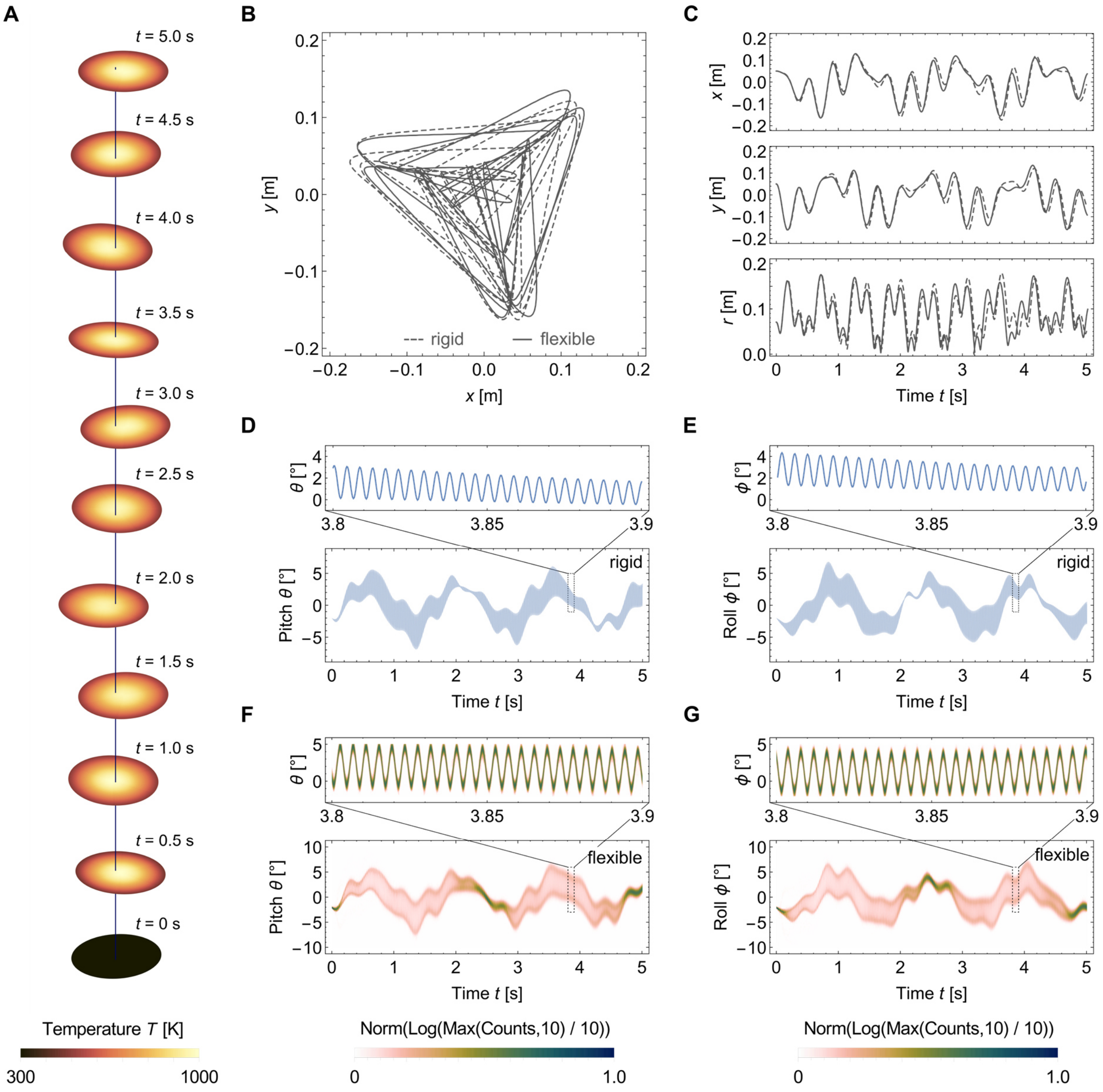

**Figure 6.** Acceleration dynamics of a flexible and a rigid spinning lightsail based on the same composite metagrating pattern. Lightsails are initially offset by $x = y = 50$ mm from the beam center and rotated by $\theta = \phi = -2°$. (A) Snapshots of the beam-riding flexible lightsail's position, angular orientation, temperature and shape at different times. *(B)* Lightsail trajectory throughout the 5 s simulation duration. *(C)* Lightsail *x*- and *y*-position and radial distance *r* from the beam center versus time, exhibiting bounded and oscillation around the equilibrium at $x$, $y = 0$. *(D), (E)* Evolution of pitch $\theta$ and roll $\phi$, respectively, of the rigid lightsail versus time, showing multi-frequency oscillation around the equilibrium at $\theta$, $\phi = 0°$. *(F), (G)* Distribution of $\theta$ and $\phi$ angles, respectively, of all mesh elements comprising the flexible lightsail versus time, showing both bounded oscillations and limited angular spread, with minor shape distortion observed through the range of surface tilt angles at any given time. For *(D)* – *(G)*, insets show fast-frequency oscillations within a reduced time window (0.1 s). An animation of this simulation is available as Supplementary Video 3

(Fig. 6B). During the entire 5 s duration of simulated propulsion, the lightsails remain within 180 cm of the beam center, as they traverse triangle-like trajectories in the $x$–$y$ plane. Comparing the trajectory of the flexible lightsail to that of the identically patterned *rigid* version, both exhibit similar behavior consistent with marginal stability. Plotting the oscillatory displacement of the lightsail centers-of-mass along $x$, $y$ and the radial distance $r$ versus time (Fig. 6C) better reveals the slight deviations in trajectory. In the beginning, both flexible and rigid lightsails follow almost indiscernible trajectories. After 0.8 seconds, differences in $x$ and $y$ become more visible, but do not grow continuously over the studied time duration, ruling out the accumulation of numerical errors due to insufficiently small time stepping as a possible reason. Instead, we attribute the small differences in position to the role of shape distortions in flexible lightsails and the effect of thermal expansion.

To elucidate the influence of temperature and thermal strain in the flexible lightsail simulations, we simulated propulsion under conditions of zero absorptivity and emissivity to keep the lightsail temperature constant at 300 K (Fig. S6A). The resulting trajectory is again very similar that of the flexible and the rigid lightsail but does not match either perfectly. However, a closer look reveals a closer resemblance in dynamics between the thermally inactive flexible lightsail and the rigid lightsail, which suggests that thermal effects play a bigger role than shape distortions, both of which exist due to the non-uniformity of the laser and optical pressure as well as the resulting non-uniform temperature distribution and thermal strain.
We also observe oscillatory motions with multiple frequency components for both translations and rotations. Examining the lightsail tilt angles $\theta$ and $\phi$ versus time for the rigid lightsail (Fig. 6D, 6E), we observe a fast-oscillating component at 240 Hz for both tilt angles associated with the assumed 120 Hz spin speed due to its two-fold cyclic symmetry (see insets) superimposed upon multiple slower nutation/precession frequencies. Throughout the simulation, although the pitch and roll angles grow larger than the initial tilt offset, both $\theta$ and $\phi$ remain bounded between $\pm 7°$. For the flexible lightsail tilt, we present the distribution of pitch and roll angles for all mesh triangles across the lightsail surface as normalized time-domain histograms in Fig. 6F and Fig. 6G. We can observe overwhelmingly similar and bounded rotation dynamics for the flexible lightsail, proving again the effectiveness of spin stabilization. At closer look at shorter time scales reveals subtle differences in the time evolution of pitch and roll angles, as indicated by an angular spread of tilt angles of ~1°.

The simulations provide a high-fidelity numerical approximation of the initial lightsail trajectory, stress distribution, and shape evolution, which is sufficient to characterize the general beam-riding and structural behavior of stable lightsail designs, and to definitively identify unstable designs. The specific design presented here appears marginally stable for the chosen initial conditions throughout the 5 s duration of acceleration. However, substantial deviations from this design and set of chosen parameters can produce unstable behavior. Decreasing the spin frequency from 120 Hz to 80 Hz, increasing the beam diameter from 0.4*D* to 0.5*D*, or increasing the gap between resonators by 20% for both TE and TM unit cells all result in unstable dynamics (Fig. S7), which highlights the importance of judiciously choosing the beam width, spin frequency and optical design for passive stabilization.

**Conclusions**

We have presented time-domain multiphysics simulations of flexible lightsail membranes undergoing the initial stages of acceleration toward relativistic velocities due to radiation pressure propulsion. In this work we have explored both the lightsail beam-riding stability and dynamic structural stability. Specifically, we have shown proof-of-concept examples of flexible, meter-scale lightsails, spin-stabilized to tension the lightsail, that exhibit a stable shape without any stiffening elements. We have observed that certain concave specularly reflecting lightsail shapes such as paraboloids can enable both beam-riding stability and shape stability, and have also demonstrated passively stabilized flat lightsail designs based on $Si_3N_4$ metagratings. The latter is of particular interest for experimental lightsail development, owing to the favorable mechanical strength and low optical absorption of $Si_3N_4$, and its ability to be fabricated in planar thin-film form at the wafer scale. Specifically, we have demonstrated that high-speed spin stabilization at 120 Hz is largely effective in rigidifying a flexible metagrating-based lightsail to exhibit similar dynamics compared to its rigid counterpart, while the subtle differences between flexible and rigid metagrating lightsails can be explained by both structural deformations and thermal effects.

We note that the size and average illumination intensity for the designs reported here fall below the nominal design targets proposed by the Breakthrough Starshot program for interstellar missions. Furthermore, the dimensions our metagratings causes them to be heavier than the nominal target of ~0.1 g/m$^2$. Further optimization of the metagratings and lightsail structure, potentially including the addition of other materials, will be necessary to produce a full-scale Starshot lightsail design. Our present design represents an important first step towards this goal, and the simulation tools reported here will likely be useful in achieving this goal.

Future work should be directed towards modelling the temperature dependence of optical reflectivity, absorptivity, and emissivity, in order to better understand the upper limits of achievable acceleration--a key factor in determining the viability of interstellar exploration via laser-propelled lightsails. For many materials, experimental efforts may be needed to probe high temperature properties. Other second-order effects may also be worthy of investigation, such as the effects of strain on optical properties. Our simulation approach may also be useful in addressing other challenges for interstellar lightsail development, such as payload integration and codesign of the propulsive laser system.

Despite numerous simplifications, we have addressed the most relevant physics for flexible lightsail acceleration and flight, including first-order linear elastic behavior, heat flow, and optical scattering. We have presented time-domain simulations of stable lightsail structures undergoing up to five seconds of acceleration. Future work may allow longer simulation durations, but regardless of the chosen simulation duration, it is difficult to infer absolute stability from time-domain simulations of marginally stable lightsails, so a more useful future application of our approach might be the improvement and optimization of lightsail designs. Our present lightsail patterning was selected based on parametric optimization under rigid-body Floquet theory, but the complexity of flexible lightsail dynamics suggests that a more advanced optimization approach based on numerical time-domain simulations may yield more favorable designs, particularly as increasingly complex building blocks and physical behaviors are modelled. Future refinements, such as implementing temperature-dependent optical properties, or improving numerical time-stepping with implicit and higher-order methods, may allow for studies of acceleration over a longer period. Nevertheless, study of the initial seconds of lightsail acceleration provides considerable insight into flexible lightsail design. In connection with the work reported here, we have published an open-source version of our simulation code (59) to further expand effort by the lightsail community to develop new and improved designs for interstellar propulsion, optical levitation, and long-range optical manipulation of macroscopic objects.

## Materials and Methods

Electromagnetic response of the TE and TM metagrating designs were simulated in COMSOL Multiphysics assuming periodic Floquet boundary conditions. For high-stress stoichiometric silicon nitride, we assumed a refractive index of $\mathrm{Re}(n) = 2$ and an extinction coefficient of $\mathrm{Im}(n) = 2 \times 10^{-6}$ at $\lambda = 1064$ nm (59). The TE and TM metagrating unit cells shown in Fig. 5B are defined by $w_1^{\mathrm{TE}} = 600$ nm, $w_1^{\mathrm{TM}} = 520$ nm, $w_2^{\mathrm{TE/TM}} = 200$ nm, $d^{\mathrm{TE}} = 1600$ nm, $d^{\mathrm{TM}} = 1350$ nm and an gap of 190 nm and 200 nm between resonators for the TE and TM unit cells, respectively. The resonators' height and substrate thickness were chosen to be 400 nm and 200 nm, respectively. The process of identifying these self-stabilizing unit cell designs, which was based on Floquet theory, i.e., evaluation of the absolute values of the eigenvalues of the monodromy (state transition) matrix, is described in more detail in the Supplementary Information. Except for the resonator height and substrate thickness, all geometrical parameters were varied systematically to select and compare suitable metagrating designs. By sweeping the incidence angle between ±25° for both pitch ($\theta$) and roll ($\phi$) tilt, angle-dependent optical pressures can be obtained via integration of the Maxwell Stress tensor around the respective unit cell.

We used the exported look-up-tables of optical pressures as inputs to our rigid and flexible membrane dynamics simulations. In the former case, optically induced forces and torques can be derived assuming a Gaussian beam characterized by its peak intensity $I_0$ and beam width $w$. For a given set of initial conditions (position, velocity, angular orientation, and angular frequency), the coupled equations of motion were evolved numerically using MATLAB's ode45 solver to obtain the trajectory and time-dependent displacement and tilt of propelled rigid lightsails described by their centers-of-mass. Normalized relevant quantities can be converted to real-life values by specifying $I_0$, the lightsail diameter $D$ and calculating the normalized time constant $t_0 = (mc/I_0)^{1/2}$, where $m$ is the total mass of the lightsail.

For more detailed description of the modeling and dynamical simulation of flexible curved and flat lightsails, we refer to the Supplementary Information. The MATLAB code has been made available on GitHub.

Perceptually uniform, undistorted color maps were used for Fig. 5C-D, 6A, 6F-G and S3A, S3F-G (60, 61).

## Acknowledgments

The authors acknowledge helpful discussions with James Benford, Victor Brar, Artur Davoyan, Ognjen Ilic, Phillip Jahelka, Adrien Merkt, Lior Michaeli, Cora Went, and Joeson Wong. We thank Zachary Manchester for insight and guidance with the Floquet stability analysis.

Funding was provided by the Air Force Office of Scientific Research under grant FA2386-18-1-4095 and the Breakthrough Starshot Initiative.

# Supplementary Information

# Dynamically Stable Radiation Pressure Propulsion of Flexible Lightsails for Interstellar Exploration

Ramon Gao[1], Michael D. Kelzenberg[1], and Harry A. Atwater*

California Institute of Technology, Pasadena CA 91125

**Tabulation of Material Properties**

We have collected a number of candidate material property values from the literature for the purpose of simulating the structural dynamics of lightsails. These appear as table S1 below. This is not intended as an exhaustive list or ranking of candidate materials for the interstellar lightsail, and importantly, it should be noted that the published properties of these materials can vary greatly depending on the method of fabrication, as well as the test geometry and method of measurement. Furthermore, most properties are reported based on room-temperature measurements, whereas during acceleration, lightsails will operate at elevated temperatures where material properties have been less comprehensively studied. We did not attempt to model temperature-dependent mechanical properties in the present study, although it would be straightforward to add this capability in the future. Ultimately, further characterization of the lightsail material(s), as fabricated and over their intended operating temperature range, will be required to draw conclusions about the viability of any specific lightsail design.

Aluminum and polyimide are typical materials used for solar sails; we include them as a point of comparison. Note that the stringent requirements of ultralow optical absorption preclude the use of even the most reflective of metals for the interstellar lightsail application. It also seems unlikely that polymers could be used structurally in this application, owing to their low strength and limited temperature range. Other materials offering exceptional mechanical strength such as graphene and carbon nanotubes can also likely be ruled out owing to their high optical absorption. However, there are likely a wide range of dielectrics and wide-bandgap semiconductors which may prove useful in lightsail applications, in addition to those shown in Table S1.

For materials such as crystalline silicon, $SiO_2$ and diamond, the highest recorded strengths have been achieved by small (< 50 µm diameter) filaments of high-purity materials with pristine surfaces, tested in bending over a small mandrel to further limit the stressed surface area and thus the chances of encountering a surface defect. It is uncertain if such high strengths could be achieved in a membrane geometry.

**Table S1.** Summary of published mechanical properties of candidate lightsail materials

| Material | Test geometry | Poisson ratio $\nu$ | Young's Modulus $E$ [GPa] | Tensile strength $\sigma_T$ [GPa] | Density $\rho$ [g/cm$^3$] | Comp. strength $\sigma_C$ [GPa] | Therm. cond. $\kappa$ [W m$^{-1}$ K$^{-1}$] | Linear CTE $\alpha_L$ [ppm K$^{-1}$] | Heat cap $C$ [J g$^{-1}$ K$^{-1}$] | Refs* |
|---|---|---|---|---|---|---|---|---|---|---|
| **Silicon** | Crystal filament | 0.06-0.28 | 130-190 | Up to 4.9 | 2.33 | 3.2 | 160 | 2.3-4.5 | 0.67 | (2-4) |
| (111) surf. | Crystal plane | 0.26 | 169 | 2.1 | 2.33 | | | | | (5, 6) |
| CVD poly | Thin film | 0.22 | 169 | 1.2 | 2.33 | | | | | (7) |
| **Diamond** | Crystal filament | 0.10-0.29 | 910-1250 | Up to 7.5 | 3.52 | 9-16 | 2000-2100 | 0.8-4.8 (1.0 typ.) | 0.51 | (8, 9) |
| CVD poly | Thin film | 0.2 | 1050 | 0.41 | | | | | | (10) |
| CVD nano | Thin film | 0.03 | 750 | 5.0 | 3.27 | | | | | (11) |
| CVD UNCD | Thin film | 0.20 | 460 | 1.8 | 2.92 | | | | | |
| **$SiO_2$** | | | | | | | | | | |
| Fused silica | Bulk | 0.16 | 73 | 0.054 | 2.20 | 1.14 | 1.4 | 0.57 | 0.77 | (12) |
| Quartz | Crystal plane | 0.16 | 97 | 0.165 | 2.65 | 2 | 10.7 | 7.1 | 0.71 | (13, 14) |
| Tempered glass | Thick film | 0.22 | 77 | Up to 1.0 | 2.40 | | 1.2 | 7.25 | 0.76 | (15, 16) |
| Silica fibers | Filament | | | Up to 6.0 | | | | | | (17) |
| **$Si_3N_4$** (LPCVD) | Thin film | 0.27 | 270 | 6.4 | 2.7 | 1-5 | 3 | 2.3 | 0.8 | (18-21) |
| **$MoS_2$** | | | | | | | | | | |
| Single-layer | 2D crystal plane | 0.27 | 270 | 22 | 5.02 | | 30-100 | 5 | 0.39 | (22-24) |
| Multi-layer | 2D crystal plane | | 200-240 | 21 | | | 80-110 | | | |
| **Aluminum** 7075 | Bulk | 0.33 | 72 | 0.50 | 2.80 | | 130 | 24 | 0.96 | |
| **Polyimide** | Thick film | 0.34 | 2.5 | 0.069 | 1.42 | | 0.12 | 20 | 1.10 | (25) |

*Note: See the SI reference list at the end of this document (not the manuscript reference list).

Furthermore, crystalline materials, whether bulk or 2D, may exhibit reduced strength if used to fabricate arbitrarily curved lightsail surfaces such as spheres, cones, or paraboloids, owing to relative weakness of certain crystal planes, or the inability to perfectly join crystal surfaces at domain boundaries.

**Mesh-based simulator for flexible lightsails**

We have developed a time-domain simulation code for studying the dynamic behavior of lightsails under acceleration. This is facilitated by modelling the lightsails as a discrete mesh, wherein the nodes represent mass, inertia, temperature, and shape; the edges represent the stiffness and thermal conductivity of the material; and enclosed triangles represent the surface area through which light interacts with the lightsail. Nodes of the mesh are assigned positions along the desired surface, with their spacing chosen to yield approximately uniform edge length and aspect ratio among the triangles. An example simulation mesh for a paraboloid lightsail is plotted in Fig. S1. This code has been open-sourced at:

https://github.com/Starshot-Lightsail

Simulations begin with generation of the mesh. Currently, supported meshes must have two-dimensional topology, but can represent any three-dimensional surface so long the surface is not self-shading at any time. The provided mesh generator script provides parametric options to generate round, square, or hexagonal lightsails, with either flat, spherical, parabolic, or cone/pyramid vertical profiles. Non-round footprints can specify either smooth or faceted vertical profiling. Region and texture mapping is also performed in the mesh generator. This assigns varying mechanical and optical properties to various regions of the lightsail. Regions can be defined in either Cartesian or polar mapping schemes.

The simulation process is outlined in Fig. S2. Briefly, evolution of the shape and position of the sail is calculated iteratively in the time domain, using a fixed time step calculated to be substantially smaller than any vibrational modes of the mesh (typically, 1/20$^{th}$ to 1/10$^{th}$ of the period of the highest resonant frequency). Modeled physics include: Mechanical response based on linear elastic theory, tensile failure detection, radiative cooling, thermal conduction, thermal expansion, ray-tracing for specular surfaces, and calculation of optical forces and absorption using either fixed values of reflectance and absorption, a 1D look-up table (LUT) to represent calculations of specular behavior using the transfer matrix method, or a 2D LUT representing the angle-dependent response of nanophotonic or metagrating surfaces.

Upon mechanical or failure of the membrane, the simulation can then be terminated, or allowed to proceed to determine the margin by which the chosen conditions will exceed the material capabilities. Alternately, to enable cursory depictions of the progression of such failures, we can delete the affected elements from the ongoing simulation mesh at the moment of failure, but this is not intended to accurately model the dynamics of tensile or thermal failures. Specifically, our simulator does not model collisions between the collapsed lightsail elements, neglects beam occlusion effects for inverted shapes, and treats tensile failure simplistically; thus the fully collapsed and tattered shapes are not simulated accurately. The images of mechanical failure are included to better show the general progression of the shape instabilities. In Fig. 3 of the main text, we chose a hexagonal perimeter shape for the flat membranes to better illustrate the collapse.

The present approach cannot be used to study the scenario of polarization mismatch, which is necessary to evaluate whether our spinning lightsails are stable in non-rotating beams, or whether lightsails can self-synchronize their rotation to that of the beam during acceleration. Moreover, due the accumulation of numerical errors introduced by explicit time stepping in our code, simulations cannot be performed over indefinite timescales to definitively prove marginal stability.

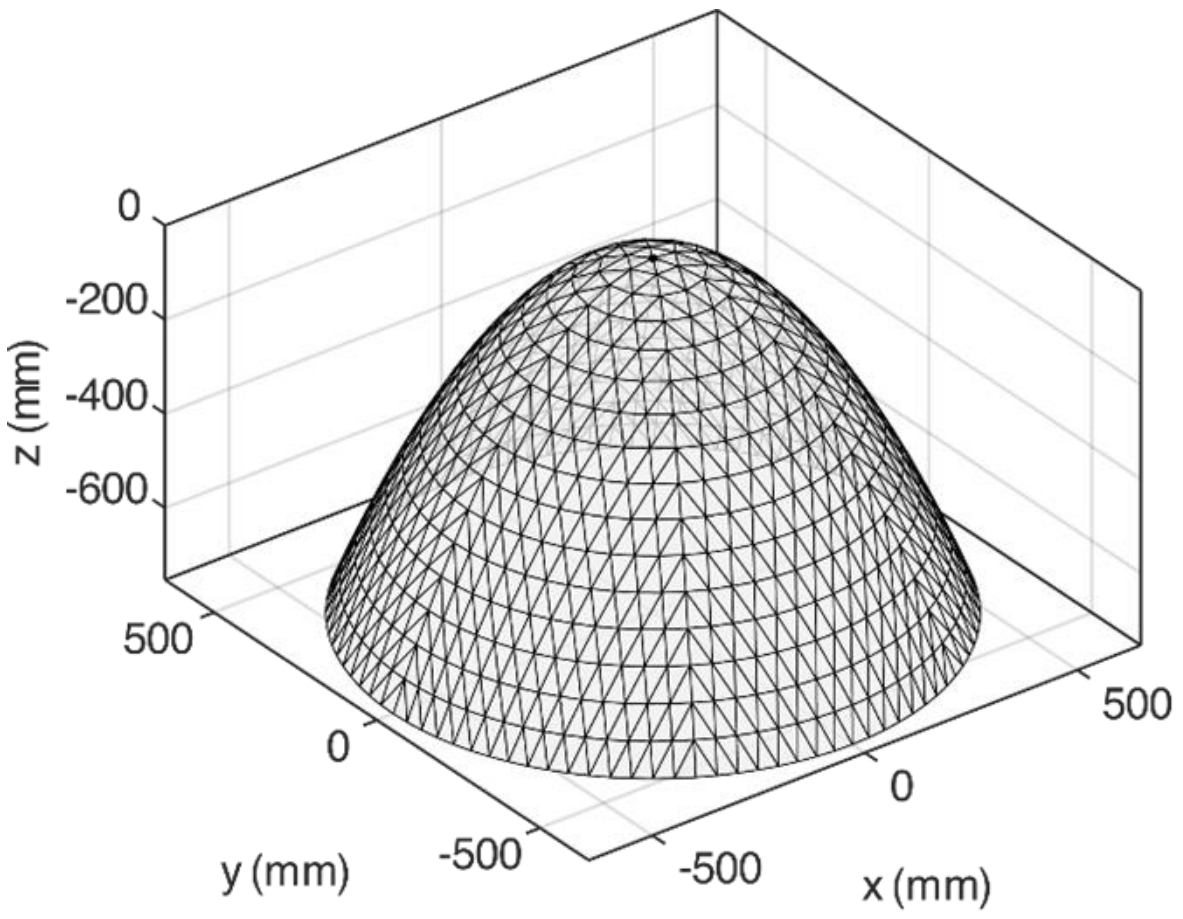

**Fig. S1.** Three-dimensional surfaces can be constructed using Delaunay triangulation to model the simulation mesh of a paraboloid lightsail.

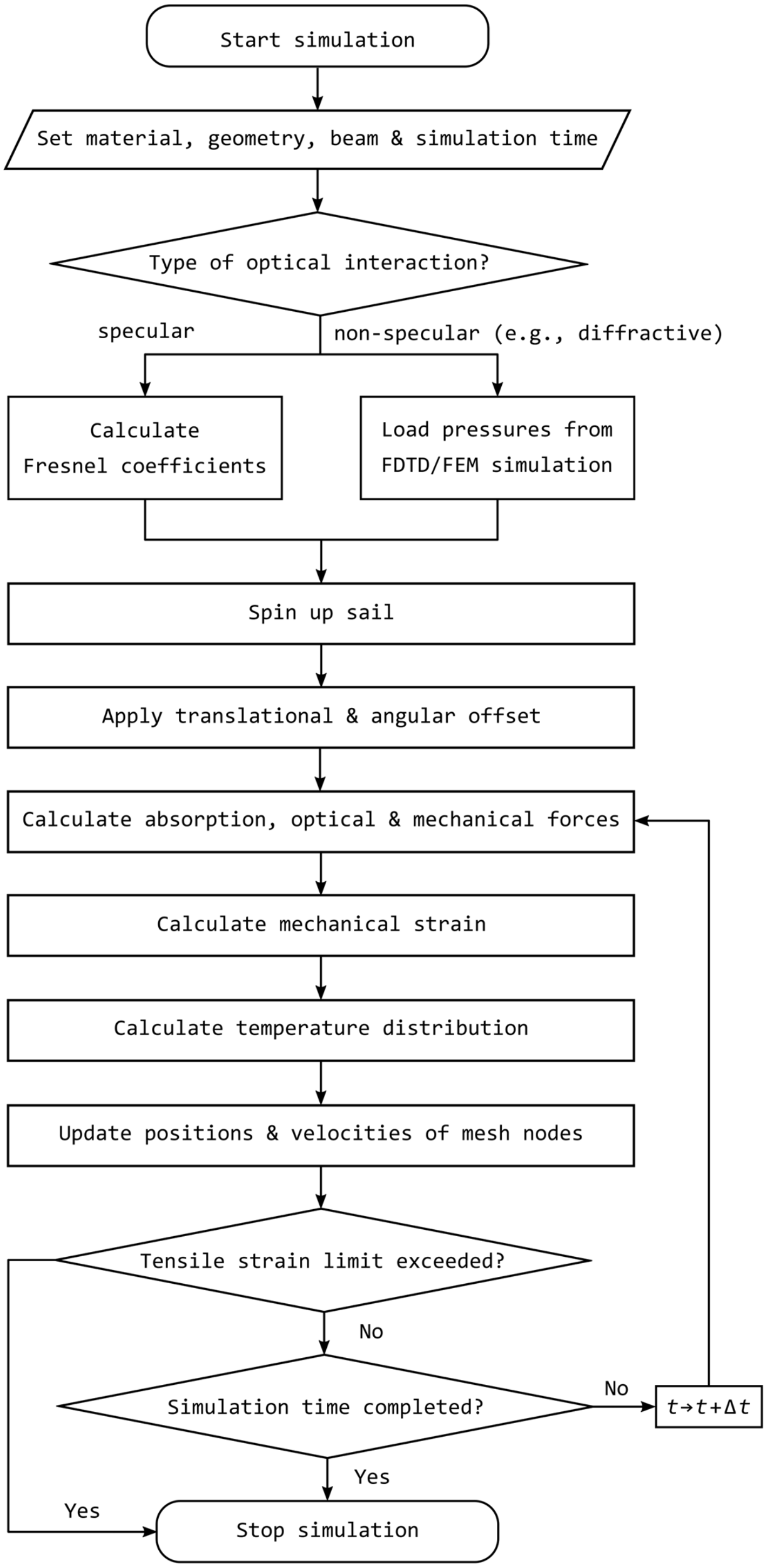

**Fig. S2.** Flow-chart diagram of flexible lightsail simulation code.

**Rigid-body lightsail dynamics**

Numerical simulations of rigid-body dynamics are based on the methods reported in (1). Motion of a meta-grating-based rigid silicon nitride lightsail is captured by its center of mass position $\boldsymbol{r} = (x, y, z)$ and angular orientation given by the Euler angles $\boldsymbol{\alpha} = (\psi, \theta, \phi)$. While $\boldsymbol{r}$ describes translational motion within the inertial lab frame $I$, $\boldsymbol{\alpha}$ denotes rotation of the lightsail or equivalently the lightsail's body frame $B$ with respect to $I$. By adopting the 1-2-3 or *x-y'-z''* rotation sequence, transformation from the inertial frame to the body frame is described the direction cosine matrix

$$\mathbf{H}_I^B(\psi, \theta, \phi) = \mathbf{R}_z(\psi)\mathbf{R}_y(\theta)\mathbf{R}_x(\phi) = \begin{pmatrix} c_1c_2 & c_1s_2s_3 + s_1c_3 & -c_1s_2c_3 + s_1s_3 \\ -s_1c_2 & -s_1s_2s_3 + c_1c_3 & s_1s_2c_3 + c_1s_3 \\ s_2 & -c_2s_3 & c_2c_3 \end{pmatrix},$$

With $c_i = \cos(\alpha_i)$, $s_i = \sin(\alpha_i)$, $\alpha_1 = \psi, \alpha_2 = \theta$ and $\alpha_3 = \phi$, and single-axis rotation matrices

$$\mathbf{R}_x(\phi) = \begin{pmatrix} 1 & 0 & 0 \\ 0 & \cos(\phi) & \sin(\phi) \\ 0 & -\sin(\phi) & \cos(\phi) \end{pmatrix}, \mathbf{R}_y(\theta) = \begin{pmatrix} \cos(\theta) & 0 & -\sin(\theta) \\ 0 & 1 & 0 \\ \sin(\theta) & 0 & \cos(\theta) \end{pmatrix},$$

$$\mathbf{R}_z(\psi) = \begin{pmatrix} \cos(\psi) & \sin(\psi) & 0 \\ -\sin(\psi) & \cos(\psi) & 0 \\ 0 & 0 & 1 \end{pmatrix}.$$

Consequently, transformation from the body frame to the inertial frame can be mathematically described by the inverse or transpose of the orthogonal direction cosine matrix

$$\mathbf{H}_B^I(\boldsymbol{\alpha}) = \left(\mathbf{H}_I^B(\boldsymbol{\alpha})\right)^T.$$

By simulating the angle-dependent optical pressures $\mathbf{p}_i(\boldsymbol{\alpha})$ on each of the four regions or sectors of the lightsail in the body frame, we can express the optically induced forces in the inertial frame, assuming Gaussian spatial variation of the beam intensity, as

$$\mathbf{F}(\mathbf{r}, \boldsymbol{\alpha}) = \mathbf{H}_B^I(\boldsymbol{\alpha}) \sum_i \mathbf{F}_i(\mathbf{r}, \boldsymbol{\alpha}),$$

$$\mathbf{F}_i(\mathbf{r}, \boldsymbol{\alpha}) = \eta \iint r d\phi dr\, \mathbf{p}_i(\boldsymbol{\alpha}) \frac{I_0}{c} \exp\left( -\frac{2}{w^2} \left\| \begin{bmatrix} x \\ y \\ 0 \end{bmatrix} + \mathbf{H}_B^I(\boldsymbol{\alpha}) \begin{bmatrix} r\cos(\varphi) \\ r\sin(\varphi) \\ 0 \end{bmatrix} \right\|^2 \right),$$

With the projected area $\eta = \cos(\theta)\cos(\phi)$, and where $i = 1, \ldots, 4$ denotes respective lightsail region with each integration limits given by

$$r_{1,2,3,4} \in \left[0, \frac{D}{2}\right], \varphi_1 \in \left[-\frac{\pi}{6}, \frac{\pi}{6}\right), \varphi_2 \in \left[\frac{\pi}{6}, \frac{5\pi}{6}\right), \varphi_3 \in \left[\frac{5\pi}{6}, \frac{7\pi}{6}\right), \varphi_4 \in \left[\frac{7\pi}{6}, \frac{11\pi}{6}\right).$$

Due to the mirror-symmetric nature of our design, we note that only regions 1 and 3 with the transverse-electric (TE) and transverse-magnetic (TM) metagrating unit cell design, respectively, need to be simulated in COMSOL for their optical pressures. Pressures on regions 2 and 4 can then related to $\mathbf{p}_{1,3}(\boldsymbol{\alpha})$ as

$$\mathbf{p}_{3,4}(\boldsymbol{\alpha}) = \begin{bmatrix} -1 & 0 & 0 \\ 0 & 1 & 0 \\ 0 & 0 & 1 \end{bmatrix} \mathbf{p}_{1,2}(-\boldsymbol{\alpha}).$$

Moreover, with regions 1 and 3 operating predominantly for TE polarization, and regions 3 and 4 operating predominantly for TM polarization, we can approximate $\hat{\mathbf{y}} \cdot \mathbf{p}_{1,3} = 0,\ \hat{\mathbf{x}} \cdot \mathbf{p}_{2,4} = 0$, i.e., regions 1 and 3 do not provide pressures in *y* direction in the body frame, whereas the same holds true for regions 2 and 4 with pressures in *x* direction in the body frame.

Optically induced torques can be similarly expressed as

$$\boldsymbol{\tau}(\mathbf{r}, \boldsymbol{\alpha}) = \sum_i \boldsymbol{\tau}_i(\mathbf{r}, \boldsymbol{\alpha}),$$

$$\boldsymbol{\tau}_i(\mathbf{r}, \boldsymbol{\alpha}) = \eta \iint r d\phi dr \begin{bmatrix} r\cos(\varphi) \\ r\sin(\varphi) \\ 0 \end{bmatrix} \times \mathbf{p}_i(\boldsymbol{\alpha}) \frac{I_0}{c} \exp\left( -\frac{2}{w^2} \left\| \begin{bmatrix} x \\ y \\ 0 \end{bmatrix} + \mathbf{H}_B^I(\boldsymbol{\alpha}) \begin{bmatrix} r\cos(\varphi) \\ r\sin(\varphi) \\ 0 \end{bmatrix} \right\|^2 \right).$$

The total force $\mathbf{F}(\mathbf{r}, \boldsymbol{\alpha})$ and torque $\boldsymbol{\tau}_i(\mathbf{r}, \boldsymbol{\alpha})$ serve as inputs to the twelve equations of motion that fully describe kinematics and dynamics of a rigid lightsail given by

$$\dot{\mathbf{r}} = \boldsymbol{v},$$

$$\dot{\boldsymbol{v}} \approx \frac{1}{m} \left( \mathbf{H}_I^B(\boldsymbol{\alpha}) \right)^T \boldsymbol{F}(\boldsymbol{r}, \boldsymbol{\alpha}),$$

$$\dot{\boldsymbol{\alpha}} = \begin{bmatrix} \dot{\psi} \\ \dot{\theta} \\ \dot{\phi} \end{bmatrix} = \begin{bmatrix} -\cos(\psi)\tan(\theta) & \sin(\psi)\tan(\theta) & 1 \\ \sin(\psi) & \cos(\psi) & 0 \\ \cos(\psi)\sec(\theta) & -\sin(\psi)\sec(\theta) & 0 \end{bmatrix} \begin{bmatrix} \omega_x \\ \omega_y \\ \omega_z \end{bmatrix} = \mathbf{L}_B^I \boldsymbol{\omega},$$

$$I_x \dot{\omega}_x = \left( I_y - I_z \right) \omega_y \omega_z + \tau_x,$$

$$I_y \dot{\omega}_y = (I_z - I_x) \omega_x \omega_z + \tau_y,$$

$$I_z \dot{\omega}_z = \left( I_x - I_y \right) \omega_x \omega_y + \tau_z.$$

Where the beam is assumed to be constant along *z*, the gravitational term is omitted due propulsion in the radiation-pressure dominated regime, and the Euler angle rates are related to the angular velocity $\boldsymbol{\omega}$ by the orthogonal matrix $\mathbf{L}_B^I$.

The principal moments of inertia $I_x, I_y, I_z$ for our round lightsail design can be derived as follows:

$$I_x = \chi^{\mathrm{TE}} \left( \int_{-\pi/6}^{\pi/6} \int_0^{D/2} r^2 \sin^2(\varphi)\, r dr d\varphi + \int_{5\pi/6}^{7\pi/6} \int_0^{D/2} r^2 \sin^2(\varphi)\, r dr d\varphi \right) + \chi^{\mathrm{TM}} \left( \int_{\pi/6}^{5\pi/6} \int_0^{D/2} r^2 \sin^2(\varphi)\, r dr d\varphi + \chi_b \int_{7\pi/6}^{11\pi/6} \int_0^{D/2} r^2 \sin^2(\varphi)\, r dr d\varphi \right),$$

$$I_y = \chi^{\mathrm{TE}} \left( \int_{-\pi/6}^{\pi/6} \int_0^{D/2} r^2 \cos^2(\varphi)\, r dr d\varphi + \int_{5\pi/6}^{7\pi/6} \int_0^{D/2} r^2 \cos^2(\varphi)\, r dr d\varphi \right) + \chi^{\mathrm{TM}} \left( \int_{\pi/6}^{5\pi/6} \int_0^{D/2} r^2 \cos^2(\varphi)\, r dr d\varphi + \chi_b \int_{7\pi/6}^{11\pi/6} \int_0^{D/2} r^2 \cos^2(\varphi)\, r dr d\varphi \right),$$

$$I_z = I_x + I_y,$$

Where $\chi^{\mathrm{TE}}$ and $\chi^{\mathrm{TM}}$ are the mass per unit area of the TE and TM region given by

$$\chi^i = \rho\left(t + h\frac{w_1^i + w_2^i}{d^i}\right), i \in \{\mathrm{TE}, \mathrm{TM}\},$$

With $\rho = 2700\ \mathrm{kg/m^3}$ being the density of silicon nitride, and the TE and TM unit cells being geometrically defined by $w_1^{\mathrm{TE}} = 600$ nm, $w_1^{\mathrm{TM}} = 520$ nm, $w_2^{\mathrm{TE/TM}} = 200$ nm, $d^{\mathrm{TE}} = 1600$ nm, $d^{\mathrm{TM}} = 1350$ nm. Hence, numerical values for $\chi^{\mathrm{TE}}$ and $\chi^{\mathrm{TM}}$ are given by $1.08 \times 10^{-3}\ \mathrm{kg/m^2}$ and $1.12 \times 10^{-3}\ \mathrm{kg/m^2}$, respectively.

Knowing the mass per unit area of the TE and TM region, the total mass of the lightsail can be calculated as

$$m = \pi\left(\frac{\chi_a}{3} + \frac{2\chi_b}{3}\right)\left(\frac{D}{2}\right)^2 \approx 0.867\ \mathrm{g},$$

For a meter-sized round lightsail, i.e., $D = 1$ m.

Following suit of (1), we can express all equations and quantities in terms of normalized units, starting with unitless lengths $\mathbf{r}' = \mathbf{r}/D$ and unitless time $t' = t/t_0$ with

$$t_0 = \sqrt{\frac{mc_0}{I_0 D}},$$

Where $c_0$ is the speed of light and $I_0$ the peak intensity of the incident Gaussian beam, which in our case was chosen to be $I_0 = 1\ \mathrm{GW/m^2}$. As a result, the spinning frequency $f_z$ can also be normalized via

$$\omega_z' = 2\pi f_z t_0.$$

Finally, we can also express forces and torques as unitless quantities with $\mathbf{F}' = \mathbf{F}/(I_0 D^2/c_0)$ and $\boldsymbol{\tau}' = \boldsymbol{\tau}/(I_0 D^3/c_0)$ as shown in Fig. 5 in the main text.

With the previously introduced equations of motion, we can numerically evolve the equations of motion by expressing them as a vectorial differential equation of first order:

$$\frac{d\mathbf{u}'}{dt} = \dot{\mathbf{u}}' = \mathbf{f}(\mathbf{u}'),\ \mathbf{u}' = (\mathbf{r}', \mathbf{v}', \boldsymbol{\alpha}, \boldsymbol{\omega}') = \left(x', y', z', v_x', v_y', v_z', \psi, \theta, \phi, \omega_x', \omega_y', \omega_z'\right),$$

Which can be numerically evolved using MATLAB's ode45 solver for given initial conditions $(\mathbf{r}_0', \mathbf{v}_0', \boldsymbol{\alpha}_0, \boldsymbol{\omega}_0')$. From here on, we drop the primed notation indicating that variables are dimensionless for better readability.

**Stability analysis of metagrating designs**

For rigid lightsails, linear stability analysis can be performed to predict whether a given composite metagrating design is unstable. We stress that conclusions about stability cannot be drawn alone from linear stability analysis due to the inexistence of damping terms in the equations of motion, hence, numerical evolution of the differential equations is necessary to confirm marginal stability. Moreover, we only consider translational and rotational degrees of freedom in our stability analysis, as there is no stabilization mechanism in propulsion direction. Hence, we can rewrite parts of the equations of motion as

$$\frac{d\tilde{\mathbf{u}}}{dt} = \dot{\tilde{\mathbf{u}}} = \tilde{\mathbf{f}}(\tilde{\mathbf{u}}),\ \tilde{\mathbf{u}} = \left(x, y, v_x, v_y, \psi, \theta, \phi, \omega_x, \omega_y, \omega_z\right), \tilde{\mathbf{f}} = \left(v_x, v_y, f_x, f_y, h_\psi, h_\theta, h_\phi, f_\phi, f_\theta, f_\psi\right).$$

In the case of a *non-spinning* rigid lightsail, we note that $\tilde{\mathbf{u}}_0 = \mathbf{0}$ is the trivial equilibrium of the system, which corresponds to the lightsail riding the beam. Consequently, the Jacobian is calculated as $\tilde{\mathbf{f}}'$ evaluated at $\tilde{\mathbf{u}}_0$:

$$\tilde{\mathbf{f}}'(\tilde{\mathbf{u}}_0) = \left.\begin{bmatrix} 0 & 0 & 1 & 0 & 0 & 0 & 0 & 0 & 0 & 0 \\ 0 & 0 & 0 & 1 & 0 & 0 & 0 & 0 & 0 & 0 \\ f_{xx} & f_{xy} & 0 & 0 & f_{x\psi} & f_{x\theta} & f_{x\phi} & 0 & 0 & 0 \\ f_{yx} & f_{yy} & 0 & 0 & f_{y\psi} & f_{y\theta} & f_{y\phi} & 0 & 0 & 0 \\ 0 & 0 & 0 & 0 & h_{\psi\psi} & h_{\psi\theta} & h_{\psi\phi} & h_{\psi\omega_x} & h_{\psi\omega_y} & h_{\psi\omega_z} \\ 0 & 0 & 0 & 0 & h_{\theta\psi} & h_{\theta\theta} & h_{\theta\phi} & h_{\theta\omega_x} & h_{\theta\omega_y} & h_{\theta\omega_z} \\ 0 & 0 & 0 & 0 & h_{\phi\psi} & h_{\phi\theta} & h_{\phi\phi} & h_{\phi\omega_x} & h_{\phi\omega_y} & h_{\phi\omega_z} \\ f_{\phi x} & f_{\phi y} & 0 & 0 & f_{\phi\psi} & f_{\phi\theta} & f_{\phi\phi} & f_{\phi\omega_x} & f_{\phi\omega_y} & f_{\phi\omega_z} \\ f_{\theta x} & f_{\theta y} & 0 & 0 & f_{\theta\psi} & f_{\theta\theta} & f_{\theta\phi} & f_{\theta\omega_x} & f_{\theta\omega_y} & f_{\theta\omega_z} \\ f_{\psi x} & f_{\psi y} & 0 & 0 & f_{\psi\psi} & f_{\psi\theta} & f_{\psi\phi} & f_{\psi\omega_x} & f_{\psi\omega_y} & f_{\psi\omega_z} \end{bmatrix}\right|_{\tilde{\mathbf{u}}_0},$$

Where we adopted the following notation for partial derivatives, $f_{ij} = \partial f_i / \partial j$.

In our case with only pitch- and roll-restoring behavior and translational stability, many of the matrix elements are either zero, very small and thus approximately zero, or can be calculated analytically, leaving us with a Jacobian $\hat{\mathbf{f}}'$ matrix of full rank that has a reduced dimension given by

$$\hat{\mathbf{f}}'(\tilde{\mathbf{u}}_0) \approx \begin{bmatrix} 0 & 0 & 1 & 0 & 0 & 0 & 0 & 0 \\ 0 & 0 & 0 & 1 & 0 & 0 & 0 & 0 \\ f_{xx}|_{\tilde{\mathbf{u}}_0} & 0 & 0 & 0 & f_{x\theta}|_{\tilde{\mathbf{u}}_0} & 0 & 0 & 0 \\ 0 & f_{yy}\big|_{\tilde{\mathbf{u}}_0} & 0 & 0 & 0 & f_{y\phi}\big|_{\tilde{\mathbf{u}}_0} & 0 & 0 \\ 0 & 0 & 0 & 0 & 0 & 0 & 0 & 1 \\ 0 & 0 & 0 & 0 & 0 & 0 & 1 & 0 \\ 0 & f_{\phi y}\big|_{\tilde{\mathbf{u}}_0} & 0 & 0 & 0 & f_{\phi\phi}\big|_{\tilde{\mathbf{u}}_0} & 0 & 0 \\ f_{\theta x}|_{\tilde{\mathbf{u}}_0} & 0 & 0 & 0 & f_{\theta\theta}|_{\tilde{\mathbf{u}}_0} & 0 & 0 & 0 \end{bmatrix}.$$

By numerically evaluating the remaining nonzero matrix elements of the Jacobian matrix, the presence of real parts in any of its eigenvalues indicates exponential growth of the respective solution to the equations of motion and thus instability of the laser-propelled system. Due to the lack of damping terms in the system's equations of motion, eigenvalues with real parts will always come in pairs of positive and negative real part.

The case of *spinning* rigid lightsails requires a more careful stability analysis, where the absolute values of the complex eigenvalues of the monodromy matrix, which can be obtained from numerical integration involving the system's Jacobian matrix, determine whether spinning lightsails are stable or not. Importantly, $\omega_z$ is no longer assumed to be zero (or close to zero), but takes on a finite value, i.e., $2\pi$ times our desired spinning frequency, instead. Similarly, the yaw angle $\psi = \psi(t)$ will vary between 0 and $2\pi$ during a period of $2\pi/\omega_z$ and thus be time-dependent. To underline these differences, we evaluate $\tilde{\mathbf{f}}'(\tilde{\mathbf{u}}_0)$ with $\omega_z$ = constant and $\psi = \psi(t)$ to be

$$\tilde{\mathbf{f}}'(\tilde{\mathbf{u}}_0) = \begin{bmatrix} 0 & 0 & 1 & 0 & 0 & 0 & 0 & 0 & 0 & 0 \\ 0 & 0 & 0 & 1 & 0 & 0 & 0 & 0 & 0 & 0 \\ f_{xx}|_{\tilde{\mathbf{u}}_0} & f_{xy}|_{\tilde{\mathbf{u}}_0} & 0 & 0 & f_{x\psi}|_{\tilde{\mathbf{u}}_0} & f_{x\theta}|_{\tilde{\mathbf{u}}_0} & f_{x\phi}|_{\tilde{\mathbf{u}}_0} & 0 & 0 & 0 \\ f_{yx}|_{\tilde{\mathbf{u}}_0} & f_{yy}|_{\tilde{\mathbf{u}}_0} & 0 & 0 & f_{y\psi}|_{\tilde{\mathbf{u}}_0} & f_{y\theta}|_{\tilde{\mathbf{u}}_0} & f_{y\phi}|_{\tilde{\mathbf{u}}_0} & 0 & 0 & 0 \\ 0 & 0 & 0 & 0 & 0 & 0 & 0 & 0 & 0 & 1 \\ 0 & 0 & 0 & 0 & 0 & 0 & 0 & \sin(\psi(t)) & \cos(\psi(t)) & 0 \\ 0 & 0 & 0 & 0 & 0 & 0 & 0 & \cos(\psi(t)) & -\sin(\psi(t)) & 0 \\ f_{\phi x}|_{\tilde{\mathbf{u}}_0} & f_{\phi y}|_{\tilde{\mathbf{u}}_0} & 0 & 0 & f_{\phi\psi}|_{\tilde{\mathbf{u}}_0} & f_{\phi\theta}|_{\tilde{\mathbf{u}}_0} & f_{\phi\phi}|_{\tilde{\mathbf{u}}_0} & 0 & -\omega_z & 0 \\ f_{\theta x}|_{\tilde{\mathbf{u}}_0} & f_{\theta y}|_{\tilde{\mathbf{u}}_0} & 0 & 0 & f_{\theta\psi}|_{\tilde{\mathbf{u}}_0} & f_{\theta\theta}|_{\tilde{\mathbf{u}}_0} & f_{\theta\phi}|_{\tilde{\mathbf{u}}_0} & \omega_z & 0 & 0 \\ f_{\psi x}|_{\tilde{\mathbf{u}}_0} & f_{\psi y}|_{\tilde{\mathbf{u}}_0} & 0 & 0 & f_{\psi\psi}|_{\tilde{\mathbf{u}}_0} & f_{\psi\theta}|_{\tilde{\mathbf{u}}_0} & f_{\psi\phi}|_{\tilde{\mathbf{u}}_0} & 0 & 0 & 0 \end{bmatrix}.$$

To further simplify, we remind ourselves that $\tilde{\mathbf{f}}(\tilde{\mathbf{u}})$ can be linearly expanded around the "equilibrium" $\tilde{\mathbf{u}}_0 = (0, 0, 0, 0, \psi(t), 0, 0, 0, 0, \omega_z)$ as

$$\dot{\tilde{\mathbf{u}}} = \tilde{\mathbf{f}}(\tilde{\mathbf{u}}) \approx \tilde{\mathbf{f}}(\tilde{\mathbf{u}}_0) + \tilde{\mathbf{f}}'(\tilde{\mathbf{u}}_0)(\tilde{\mathbf{u}} - \tilde{\mathbf{u}}_0),$$

Noting that

$$\tilde{\mathbf{f}}(\tilde{\mathbf{u}}_0) = \begin{bmatrix} 0 \\ 0 \\ 0 \\ 0 \\ \omega_z \\ 0 \\ 0 \\ 0 \\ 0 \\ 0 \end{bmatrix},$$

Which means that $\tilde{\mathbf{u}}_0$ is not a true equilibrium. Nevertheless, evaluating the second term on the right-hand side of the Taylor-expanded equation above yields

$$\tilde{\mathbf{f}}'(\tilde{\mathbf{u}}_0)(\tilde{\mathbf{u}} - \tilde{\mathbf{u}}_0) = \tilde{\mathbf{f}}'(\tilde{\mathbf{u}}_0)\tilde{\mathbf{u}} - \tilde{\mathbf{f}}'(\tilde{\mathbf{u}}_0)\tilde{\mathbf{u}}_0 = \tilde{\mathbf{f}}'(\tilde{\mathbf{u}}_0)\tilde{\mathbf{u}} - \begin{bmatrix} 0 \\ 0 \\ f_{x\psi}|_{\tilde{\mathbf{u}}_0}\psi(t) \\ f_{y\psi}|_{\tilde{\mathbf{u}}_0}\psi(t) \\ \omega_z \\ 0 \\ 0 \\ f_{\phi\psi}|_{\tilde{\mathbf{u}}_0}\psi(t) \\ f_{\theta\psi}|_{\tilde{\mathbf{u}}_0}\psi(t) \\ f_{\psi\psi}|_{\tilde{\mathbf{u}}_0}\psi(t) \end{bmatrix},$$

Such that

$$\tilde{\mathbf{f}}(\tilde{\mathbf{u}}) \approx \tilde{\mathbf{f}}(\tilde{\mathbf{u}}_0) + \tilde{\mathbf{f}}'(\tilde{\mathbf{u}}_0)(\tilde{\mathbf{u}} - \tilde{\mathbf{u}}_0) = \tilde{\mathbf{f}}'(\tilde{\mathbf{u}}_0)\tilde{\mathbf{u}} - \begin{bmatrix} 0 \\ 0 \\ f_{x\psi}|_{\tilde{\mathbf{u}}_0}\psi(t) \\ f_{y\psi}|_{\tilde{\mathbf{u}}_0}\psi(t) \\ 0 \\ 0 \\ 0 \\ f_{\phi\psi}|_{\tilde{\mathbf{u}}_0}\psi(t) \\ f_{\theta\psi}|_{\tilde{\mathbf{u}}_0}\psi(t) \\ f_{\psi\psi}|_{\tilde{\mathbf{u}}_0}\psi(t) \end{bmatrix}$$

$$= \begin{bmatrix} 0 & 0 & 1 & 0 & 0 & 0 & 0 & 0 & 0 & 0 \\ 0 & 0 & 0 & 1 & 0 & 0 & 0 & 0 & 0 & 0 \\ f_{xx}|_{\tilde{\mathbf{u}}_0} & f_{xy}|_{\tilde{\mathbf{u}}_0} & 0 & 0 & 0 & f_{x\theta}|_{\tilde{\mathbf{u}}_0} & f_{x\phi}|_{\tilde{\mathbf{u}}_0} & 0 & 0 & 0 \\ f_{yx}|_{\tilde{\mathbf{u}}_0} & f_{yy}|_{\tilde{\mathbf{u}}_0} & 0 & 0 & 0 & f_{y\theta}|_{\tilde{\mathbf{u}}_0} & f_{y\phi}|_{\tilde{\mathbf{u}}_0} & 0 & 0 & 0 \\ 0 & 0 & 0 & 0 & 0 & 0 & 0 & 0 & 0 & 1 \\ 0 & 0 & 0 & 0 & 0 & 0 & 0 & \sin(\psi(t)) & \cos(\psi(t)) & 0 \\ 0 & 0 & 0 & 0 & 0 & 0 & 0 & \cos(\psi(t)) & -\sin(\psi(t)) & 0 \\ f_{\phi x}|_{\tilde{\mathbf{u}}_0} & f_{\phi y}|_{\tilde{\mathbf{u}}_0} & 0 & 0 & 0 & f_{\phi\theta}|_{\tilde{\mathbf{u}}_0} & f_{\phi\phi}|_{\tilde{\mathbf{u}}_0} & 0 & -\omega_z & 0 \\ f_{\theta x}|_{\tilde{\mathbf{u}}_0} & f_{\theta y}|_{\tilde{\mathbf{u}}_0} & 0 & 0 & 0 & f_{\theta\theta}|_{\tilde{\mathbf{u}}_0} & f_{\theta\phi}|_{\tilde{\mathbf{u}}_0} & \omega_z & 0 & 0 \\ f_{\psi x}|_{\tilde{\mathbf{u}}_0} & f_{\psi y}|_{\tilde{\mathbf{u}}_0} & 0 & 0 & 0 & f_{\psi\theta}|_{\tilde{\mathbf{u}}_0} & f_{\psi\phi}|_{\tilde{\mathbf{u}}_0} & 0 & 0 & 0 \end{bmatrix} \tilde{\mathbf{u}}.$$

Noting that $f_{\psi x}|_{\tilde{\mathbf{u}}_0} = f_{\psi y}|_{\tilde{\mathbf{u}}_0} = 0$ and $f_{\psi\theta}|_{\tilde{\mathbf{u}}_0} = f_{\psi\phi}|_{\tilde{\mathbf{u}}_0} = 0$ due to the absence of a yaw-restoring mechanism, we get

$$\tilde{\mathbf{f}}(\tilde{\mathbf{u}}) \approx \begin{bmatrix} 0 & 0 & 1 & 0 & 0 & 0 & 0 & 0 & 0 & 0 \\ 0 & 0 & 0 & 1 & 0 & 0 & 0 & 0 & 0 & 0 \\ f_{xx}|_{\tilde{\mathbf{u}}_0} & f_{xy}|_{\tilde{\mathbf{u}}_0} & 0 & 0 & 0 & f_{x\theta}|_{\tilde{\mathbf{u}}_0} & f_{x\phi}|_{\tilde{\mathbf{u}}_0} & 0 & 0 & 0 \\ f_{yx}|_{\tilde{\mathbf{u}}_0} & f_{yy}|_{\tilde{\mathbf{u}}_0} & 0 & 0 & 0 & f_{y\theta}|_{\tilde{\mathbf{u}}_0} & f_{y\phi}|_{\tilde{\mathbf{u}}_0} & 0 & 0 & 0 \\ 0 & 0 & 0 & 0 & 0 & 0 & 0 & 0 & 0 & 1 \\ 0 & 0 & 0 & 0 & 0 & 0 & 0 & \sin(\psi(t)) & \cos(\psi(t)) & 0 \\ 0 & 0 & 0 & 0 & 0 & 0 & 0 & \cos(\psi(t)) & -\sin(\psi(t)) & 0 \\ f_{\phi x}|_{\tilde{\mathbf{u}}_0} & f_{\phi y}|_{\tilde{\mathbf{u}}_0} & 0 & 0 & 0 & f_{\phi\theta}|_{\tilde{\mathbf{u}}_0} & f_{\phi\phi}|_{\tilde{\mathbf{u}}_0} & 0 & -\omega_z & 0 \\ f_{\theta x}|_{\tilde{\mathbf{u}}_0} & f_{\theta y}|_{\tilde{\mathbf{u}}_0} & 0 & 0 & 0 & f_{\theta\theta}|_{\tilde{\mathbf{u}}_0} & f_{\theta\phi}|_{\tilde{\mathbf{u}}_0} & \omega_z & 0 & 0 \\ 0 & 0 & 0 & 0 & 0 & 0 & 0 & 0 & 0 & 0 \end{bmatrix} \begin{bmatrix} x \\ y \\ v_x \\ v_y \\ \psi(t) \\ \theta \\ \phi \\ \omega_x \\ \omega_y \\ \omega_z \end{bmatrix}.$$

From this, it follows that

$$\dot{\omega}_z(t) \approx 0 \rightarrow \omega_z = \text{constant},$$

$$\dot{\psi}(t) \approx \omega_z(t) = \omega_z \rightarrow \psi(t) = \omega_z t,$$

And we can reduce the ten-dimensional vectorial differential equation by two more dimensions to

$$\hat{\mathbf{f}}(\hat{\mathbf{u}}) \approx \hat{\mathbf{J}}(\hat{\mathbf{u}}_0, t)\hat{\mathbf{u}},$$

With

$$\hat{\mathbf{J}}(\hat{\mathbf{u}}_0, t)\hat{\mathbf{u}} = \begin{bmatrix} 0 & 0 & 1 & 0 & 0 & 0 & 0 & 0 \\ 0 & 0 & 0 & 1 & 0 & 0 & 0 & 0 \\ f_{xx}|_{\hat{\mathbf{u}}_0} & f_{xy}\big|_{\hat{\mathbf{u}}_0} & 0 & 0 & f_{x\theta}|_{\hat{\mathbf{u}}_0} & f_{x\phi}\big|_{\hat{\mathbf{u}}_0} & 0 & 0 \\ f_{yx}\big|_{\hat{\mathbf{u}}_0} & f_{yy}\big|_{\hat{\mathbf{u}}_0} & 0 & 0 & f_{y\theta}\big|_{\hat{\mathbf{u}}_0} & f_{y\phi}\big|_{\hat{\mathbf{u}}_0} & 0 & 0 \\ 0 & 0 & 0 & 0 & 0 & 0 & \sin(\omega_z t) & \cos(\omega_z t) \\ 0 & 0 & 0 & 0 & 0 & 0 & \cos(\omega_z t) & -\sin(\omega_z t) \\ f_{\phi x}\big|_{\hat{\mathbf{u}}_0} & f_{\phi y}\big|_{\hat{\mathbf{u}}_0} & 0 & 0 & f_{\phi\theta}\big|_{\hat{\mathbf{u}}_0} & f_{\phi\phi}\big|_{\hat{\mathbf{u}}_0} & 0 & -\omega_z \\ f_{\theta x}|_{\hat{\mathbf{u}}_0} & f_{\theta y}\big|_{\hat{\mathbf{u}}_0} & 0 & 0 & f_{\theta\theta}|_{\hat{\mathbf{u}}_0} & f_{\theta\phi}\big|_{\hat{\mathbf{u}}_0} & \omega_z & 0 \end{bmatrix} \begin{bmatrix} x \\ y \\ v_x \\ v_y \\ \theta \\ \phi \\ \omega_x \\ \omega_y \end{bmatrix},$$

And

$$\hat{\mathbf{u}}_0 = \mathbf{0}.$$

To continue with the stability analysis, given that we now have a linear *time-dependent* system for spinning lightsails, we need to find the monodromy matrix, i.e., the state transition matrix after one full period $T = 2\pi/\omega_z$. To find the general state transition matrix, we need to solve for

$$\dot{\mathbf{\Phi}}(t, 0) = \hat{\mathbf{J}}(\hat{\mathbf{u}}_0, t)\mathbf{\Phi}(t, 0); \qquad \mathbf{\Phi}(0, 0) = \mathbf{I}.$$

$\mathbf{\Phi}(t = T, 0)$ can be found via numerical integration of the differential equation. Specifically, writing the state transition matrix $\mathbf{\Phi}(t, 0)$ in terms of its row vectors

$$\mathbf{\Phi}(t, 0) = \begin{bmatrix} \mathbf{a}_1^T \\ \mathbf{a}_2^T \\ \vdots \\ \mathbf{a}_7^T \\ \mathbf{a}_8^T \end{bmatrix},$$

With $\mathbf{a}_i = \{a_{i1}, a_{i2}, \dots, a_{i8}\}^T \in \mathbb{R}^8, i = 1, \dots, 8$ being an eight-dimensional column vector, we can express the differential equation above in vectorial form as

$$\begin{bmatrix} \dot{\mathbf{a}}_1 \\ \dot{\mathbf{a}}_2 \\ \vdots \\ \dot{\mathbf{a}}_7 \\ \dot{\mathbf{a}}_8 \end{bmatrix} = \begin{bmatrix} \left(\mathbf{j}_1^T \cdot \mathbf{\Phi}(t, 0)\right)^T \\ \left(\mathbf{j}_2^T \cdot \mathbf{\Phi}(t, 0)\right)^T \\ \vdots \\ \left(\mathbf{j}_2^T \cdot \mathbf{\Phi}(t, 0)\right)^T \\ \left(\mathbf{j}_2^T \cdot \mathbf{\Phi}(t, 0)\right)^T \end{bmatrix},$$

With the 64-dimensional state vector $\mathbf{s} = [a_{11}(t), \dots, a_{18}(t), a_{21}(t), \dots, a_{28}(t), \dots, a_{88}(t)]^T$.

We solved this vector differential equation of first order in MATLAB using ode45 to obtain a numerical result for $\mathbf{\Phi}(t = T, 0)$. Finally, using MATLAB's eigenvalue solver, we calculated the eight eigenvalues of $\mathbf{\Phi}(t = T, 0)$ to be

$$\lambda_{1,2} \approx 0.9859 \pm 0.1671i,$$

$$\lambda_{3,4} \approx 0.9906 \pm 0.1367i,$$

$$\lambda_{5,6} \approx 0.9995 \pm 0.0308i,$$

$$\lambda_{7,8} \approx 0.9999 \pm 0.0001i.$$

The absolute values of the complex eigenvalues, $|\lambda_i|$ for $i = 1, \dots, 8$, are then given by

$$|\lambda_i| \approx 1 \; \forall i,$$

Rounded off to four decimal places.

According to Floquet theory, with the absolute values of the eight eigenvalues being on the unit circle in the complex plane, we can conclude that the presented composite metagrating design enables marginal stability in linearized dynamics of spinning flat lightsails. However, to verify that actual trajectories with finite initial conditions are bounded, the corresponding equations of motion need to be numerically evolved.

**Setting up optical forces on meshed flexible lightsail**

Prior to simulated release and propulsion of the mesh-based lightsail, each triangular mesh element $m$ is characterized by a normal vector $\mathbf{n}_m$ and a "grating" vector $\mathbf{g}_m$ being parallel to the metagratings this particular mesh element is meant to host. After launch, both vectors will evolve according to the specific rotation of that mesh element, now described by a *rotated* normal vector $\tilde{\mathbf{n}}_m$ and a *rotated* grating vector $\tilde{\mathbf{g}}_m$ at time $t$. Assuming all vectors to be normalized to a length of 1, we can then calculate the matrix that describes rotation of mesh element $k$ as

$$\mathbf{R}_m = \begin{bmatrix} \tilde{\mathrm{g}}_{\mathrm{x},m} & \tilde{\mathrm{n}}_{\mathrm{x},m} & (\tilde{\mathbf{g}}_m \times \tilde{\mathbf{n}}_m)_{\mathrm{x}} \\ \tilde{\mathrm{g}}_{\mathrm{y},m} & \tilde{\mathrm{n}}_{\mathrm{y},m} & (\tilde{\mathbf{g}}_m \times \tilde{\mathbf{n}}_m)_{\mathrm{y}} \\ \tilde{\mathrm{g}}_{\mathrm{z},m} & \tilde{\mathrm{n}}_{\mathrm{z},m} & (\tilde{\mathbf{g}}_m \times \tilde{\mathbf{n}}_m)_{\mathrm{z}} \end{bmatrix} \begin{bmatrix} \mathrm{g}_{\mathrm{x},m} & \mathrm{g}_{\mathrm{y},m} & \mathrm{g}_{\mathrm{z},m} \\ \mathrm{n}_{\mathrm{x},m} & \mathrm{n}_{\mathrm{y},m} & \mathrm{n}_{\mathrm{z},m} \\ (\mathbf{g}_m \times \mathbf{n}_m)_{\mathrm{x}} & (\mathbf{g}_m \times \mathbf{n}_m)_{\mathrm{y}} & (\mathbf{g}_m \times \mathbf{n}_m)_{\mathrm{z}} \end{bmatrix}.$$

Knowledge of the components of $\mathbf{R}_m = \mathbf{H}_B^I(\psi_m, \theta_m, \phi_m)$ allows us to calculate the Euler angles of respective mesh element

$$\theta_m = \sin^{-1}(R_{13,m}),$$

$$\phi_m = -\tan^{-1}\left(\frac{R_{23,m}}{R_{33,m}}\right),$$

$$\psi_m = -\tan^{-1}\left(\frac{R_{12,m}}{R_{11,m}}\right),$$

Where we use MATLAB's four-quadrant inverse tangent function atan2 in our code for correct calculation of $\phi_m$ and $\psi_m$.

After looking up the tabulated pressure $\mathbf{p}_m$ on mesh element $m$ based on its calculated Euler angles $(\theta_m, \phi_m)$ (noting that our pressures do not depend on $\psi$ due to the assumption of a synchronously spinning electric field), we can determine the optically induced force on specific the mesh element in the body frame to be

$$\mathbf{F}_m^B = \cos(\theta_m)\cos(\phi_m)\, A_m \mathbf{p}_m(\theta_m, \phi_m) I_m,$$

Where $A_m$ is the area of the mesh element and $I_m$ is the discretized value of the Gaussian intensity function present on the mesh element. As before, optical forces in the inertial frame are obtained by multiplication of $\mathbf{F}_m^B$ with the direction cosine matrix as follows

$$\mathbf{F}_m = \mathbf{H}_B^I(\psi_m, \theta_m, \phi_m)\mathbf{F}_m^B.$$

Due to the interconnected nature of mesh elements, torque contributions are inherently accounted for by the collective effect of optical forces on individual mesh elements.

**Example of an initially translated metagrating-based lightsail being passive stabilized**

To further probe the stability of the reported metagrating-based lightsail design, we simulated another case where the lightsail is initially displaced, but not tilted. Specifically, we assumed an initial translation offset of $x = y = 0.05$ m as in the case discussed in the main text. Setting the initial tilt offset to zero leads to considerably smaller tilt deviations and lateral displacements throughout the flight, as is evident from the snapshots shown in Fig. S3A. An animation of this simulation is available as Supplementary Video 4. The temperature distribution and peak temperature is generally similar to that of Fig. 6A. In terms of the trajectory and translational degrees of freedom, both the rigid and the flexible sails appear to exhibit passive stability as evidenced by their similar bounded transverse motion; in fact, the trajectories are more regular and less complex than those shown in Fig. 6. Lateral displacements of up to 0.08 m are observed for both the rigid and flexible case (Fig. S3B and Fig. S3C), with more apparent, yet small deviations starting to appear after the first second of acceleration.

As for the first case, we observe multiple frequency components within the simulated trajectories and tilt angles (Fig. S3D–2G), with the most noticeable one being the again slow frequency component at 240 Hz superimposed upon slower frequencies of approximately 2.5 Hz and 0.6 Hz. The observation of displacement along $x$ and $y$ being more tightly confined can also be made for the pitch and roll angles of the rigid lightsail, as they remain bounded within ±1.3° during the simulated timespan, suggesting a lesser degree of deformation and vibration in the membrane. The temporal evolution of pitch and roll angle distributions of the flexible lightsail again follows closely $\theta$ and $\phi$ of the rigid lightsail, confirming that spin stabilization at 120 Hz is sufficiently fast enough to treat our flexible lightsail as quasi-rigid. Nevertheless, we note that a finite angular spread of pitch and roll angles of ~1° can be observed for all mesh elements constituting the flexible lightsail. Finally, due to the discretized surface of the flexible lightsail, signs of mesh elements on the perimeter experiencing larger rotations remain visible in the insets of Fig. S3F and S3G despite truncating histogram bins with only few elements (less than 10 within bins of width 0.05°).

**Temperature & strain analysis of passively stabilized flexible metagrating-based lightsails**

As mentioned in the main text, our flexible lightsail simulator stores several variables of interest for post-processing and analysis, including the peak and average temperature of the lightsail during propulsion and the maximum strain on the lightsail due to mechanical forces and thermal expansion, downsampled by a factor of 8 for memory management. Due to the underfilling beam width of $w = 0.4D$, regardless of whether the lightsail is initially only translated or also tilted, the difference between the peak, average and minimum temperature of points on the lightsail can be several hundreds of Kelvin (Fig. S4). While the center of the lightsail heats up to a peak temperature of just below 1000 K during propulsion, its perimeter or edge points experience a temperature rise of less than 200 K, the difference of which results in an average temperature in between these two extremes. Including an initial tilt to the simulated trajectories induces more variation in especially both peak (center) and minimum (edge) temperatures of the accelerated lightsail.

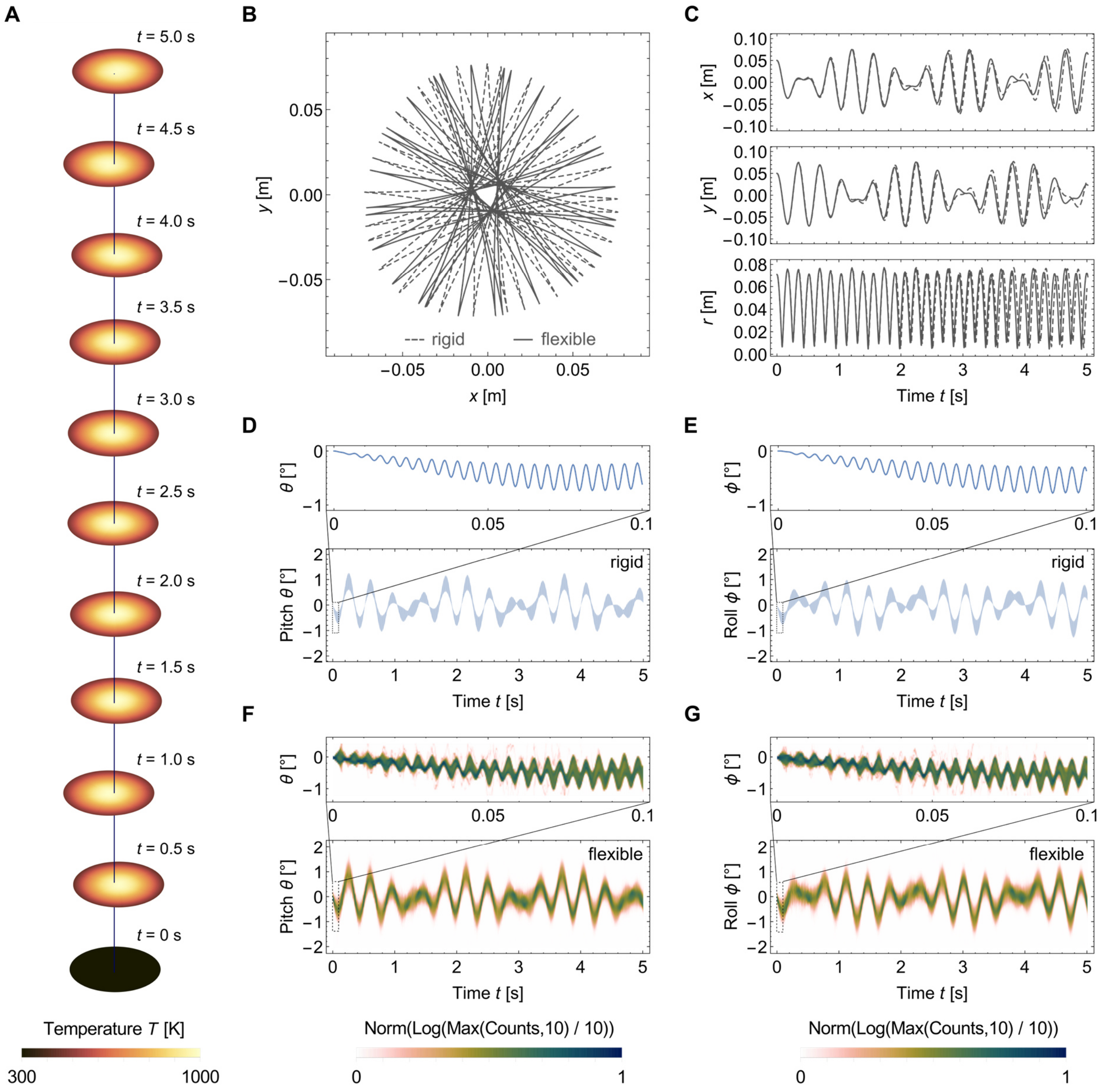

**Fig. S3.** Acceleration dynamics of a flexible and a rigid spinning composite metagrating lightsail subject to an initial translational offset. Lightsails are initially offset by $x = y = 0.05D = 50$ mm relative to the beam center**. (A)** Snapshots of the position, angular orientation, temperature and shape of the flexible lightsail at different times of stable beam-riding during propulsion. **(B)** Lateral trajectory of the flexible lightsail versus that of a rigid lightsail with the same metagrating patterning throughout the 5 s simulation duration. **(C)** Sail $x$- and $y$-position and radial distance $r$ from the beam center versus time for the flexible and rigid version of the same lightsail, exhibiting bounded and oscillation around the equilibrium at $x$, $y = 0$. **(D), (E)** Evolution of pitch $\theta$ and roll $\phi$, respectively, of the rigid lightsail versus time, showing multi-frequency oscillation around the equilibrium at $\theta$, $\phi = 0°$. **(F), (G)** Distribution of $\theta$ and $\phi$ angles, respectively, of all mesh elements comprising the flexible lightsail versus time, showing both bounded oscillations and limited angular spread, although minor shape distortion is observed via the range of surface tilt angles at

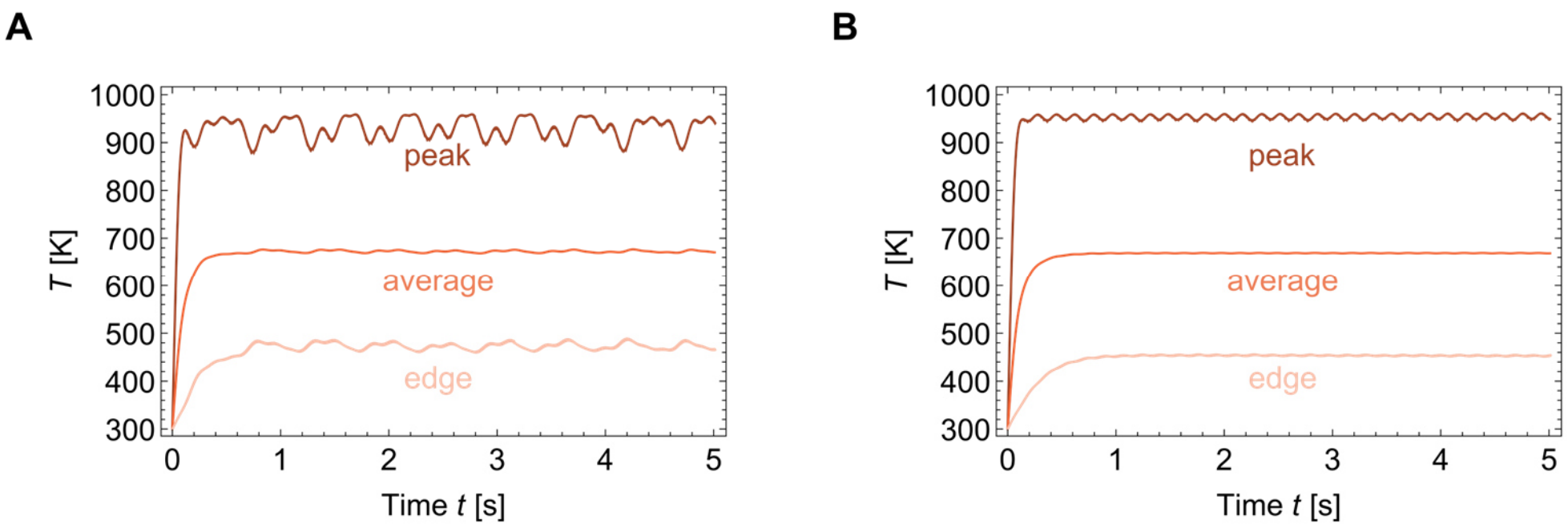

**Fig. S4.** Peak (brown) and average (orange) lightsail temperature versus time for cases discussed in the main text, where a flexible metagrating-patterned lightsail is either **(A)** being only initially translated by $x = y = 0.05$ m, or **(B)** initially translated by $x = y = 0.05$ m and initially tilted by $\theta = \phi = -2°$. Greater temperature variations are observed in (B) because the lightsail's lateral oscillations are considerably larger.

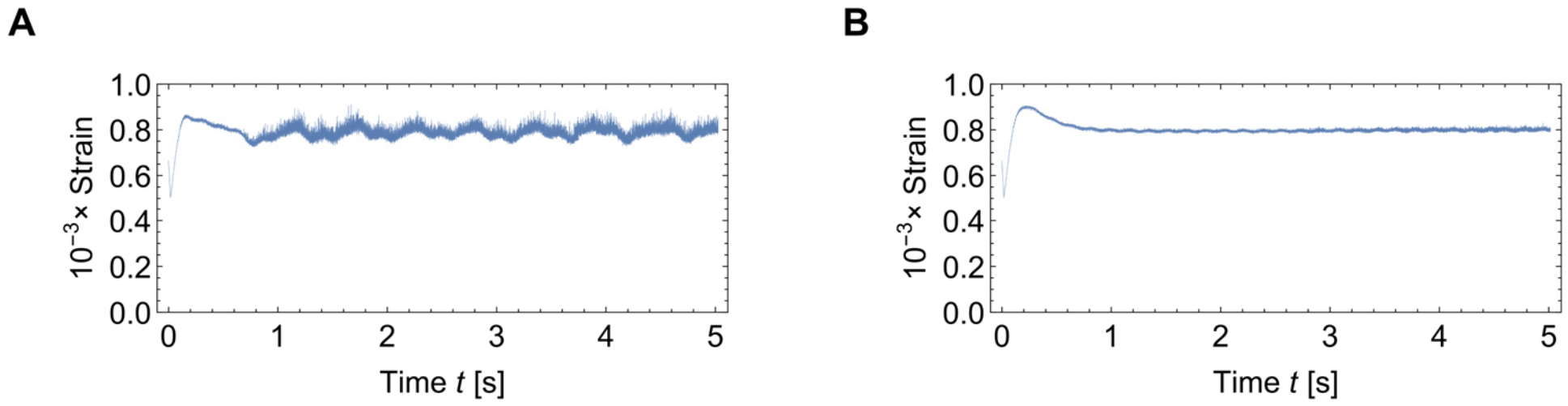

**Fig. S5.** Maximum strain on the flexible metagrating-patterned sails versus time for cases discussed in the main text, where a flexible, metagrating-patterned lightsail is either (A) being only initially translated by $x = y = 0.05$ m, or (B) initially translated by $x = y = 0.05$ m and tilted by $\theta = \phi = -2°$. Note that based on the modulus and tensile strength of $Si_3N_4$ (see Table 1), the material strain limit is ~4%; over 40x greater than the peak simulated strain.

While the maximum temperature is of interest for assessing the need to include temperature-dependent material properties and requirements for payload integration, insights on the maximum strain help to deduce how close the lightsail is to mechanical failure and breaking apart. As seen in Fig. S5, for both studied cases discussed in the main text, the maximum strain stays below 0.001, which multiplied with stoichiometric silicon nitride's Young's modulus of 270 GPa is one order of magnitude lower than its tensile strength.

**Effect of temperature on dynamics of flexible lightsails**

To study the influence of finite absorption and thus temperature rise of flexible lightsails during propulsion on their dynamics, additional simulations were run with the absorptivity set to zero. This prevents the lightsail before heating up. At the same time, to avoid the lightsail from cooling down to unrealistically low temperatures, not taking into account the temperature of space, we also set the emissivity to zero. Results for the two case studies discussed in the main text are shown in Fig. S6, comparing the trajectory of a flexible lightsail with finite absorptivity with that of a flexible lightsail with zero absorptivity and with that of a rigid lightsail with no heat transfer physics at all for initial translation of $x = y = 0.05$ m and tilt $\theta = \phi = -2°$ (Fig. S6A) and for initial translation $x = y = 0.05$ m only (Fig. S6B). We observe that while all three cases share similar dynamical behavior, the trajectory of the non-absorbing flexible lightsail is distinct from the other two. This observation becomes clearer when looking at $x$ and $y$ versus time for all three cases in Fig. S6C and Fig. S6D. The slight deviation of the non-absorbing lightsail trajectory from the absorbing lightsail trajectory hints at the combined role of temperature *and* structural flexibility on the altered dynamics of a realistic, i.e., absorbing flexible lightsail when compared to its rigid counterpart.

## Unstable cases of propelled flexible metagrating-based lightsails

The two cases discussed in the main text and in the previous section show passively stabilized dynamics of suitably designed metagrating-based lightsails with adequately high spinning frequency (120 Hz) and sufficiently small, i.e., underfilling propulsion beam width (0.4$D$). Changes to this parameter can result in unbounded and thus unstable trajectories assuming an initial translation of $x = y = 0.05$m and initial tilt of $\theta = \phi = -2°$. For example, changing only the spinning frequency by reducing it from 120 Hz to 80 Hz results in unstable dynamics for the assumed initial conditions (Fig. S7A). While Floquet theory still predicts marginal stability for our metagrating design at 80 Hz, it is the combination of a sufficiently large tilt and more structural deformation due to weaker spin-induced tensioning forces that explains the unstable behavior. The same observation of lost passive stabilization is made when increasing only the beam width instead by 25% to 0.5$D$ (Fig. S7B), possibly due to departure from the linear regime assumed in Floquet theory for the chosen finite initial tilt. Finally, the beam-riding stability depends on the restoring forces and torques produced by the chosen metagrating designs. It is likely that more optimal metagrating designs exist, but within our design space for the metagrating, a vast majority of design parameter choices produce unstable or less stable lightsails. One example is shown in Fig. S7C: In this case, increasing the gap between resonators for both the TE and TM unit cells by 20% causes the lightsail to veer off from the beam path and become unstable within the first second of flight (Fig. S7C). While all three cases were simulated to be unstable due to unbounded trajectories (Fig. S7D), the third case of metagrating resonators being spaced farther apart can be explained theoretically. Increasing the distance flips the sign of the calculated torque $\tau_x$, which causes the lightsail to lose its roll-restoring ability and thus become unstable (Fig. S7E). This conclusion is further corroborated by the fact that the complex eigenvalues of the corresponding monodromy matrix are calculated as

$$\lambda_{1,2} = 1.0666 \pm 0.043i,$$

$$\lambda_{3,4} = 0.9968 \pm 0.0805i,$$

$$\lambda_{5,6} = 0.9361 \pm 0.0377i,$$

$$\lambda_{7,8} = 0.9999 \pm 0.0001i,$$

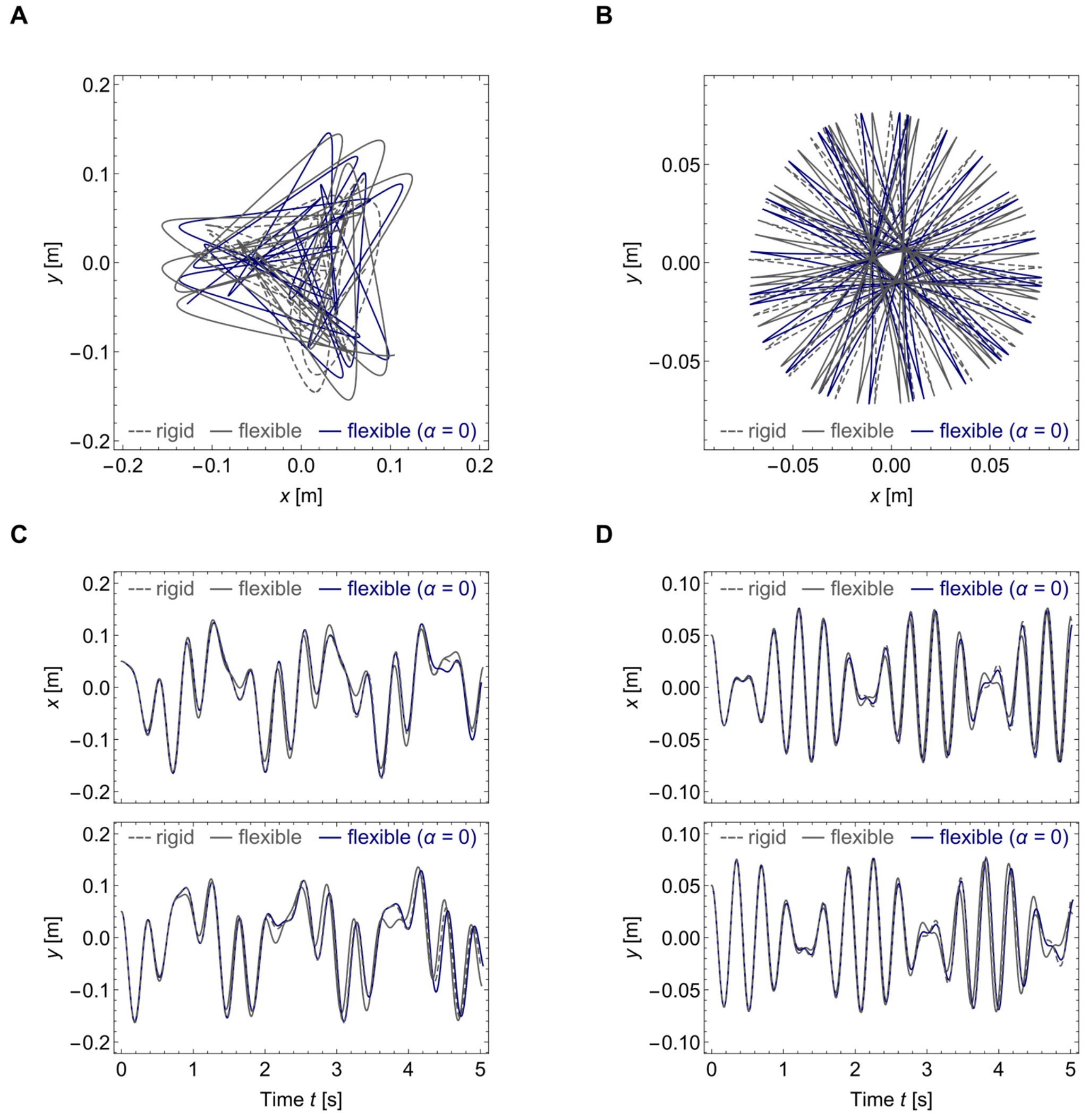

**Fig. S6.** Simulated $x$–$y$ trajectories of rigid flat lightsail (gray dashed line), flexible flat lightsail (gray line) and flexible flat lightsail with zero absorption and emission (blue line), all patterned with the presented composite metagrating design for (A) initial translation of $x = y = 0.05$ m and (B) initial translation of $x = y = 0.05$ m and initial tilt of $\theta = \phi = -2°$. The respective lightsail's $x$- and $y$-position are plotted versus time in (C) and (D) for initial translation only and both initial translation and rotation, respectively, highlighting both similarities and differences and thus the combined role of thermal expansion and shape distortions on the flexible lightsail's dynamics.

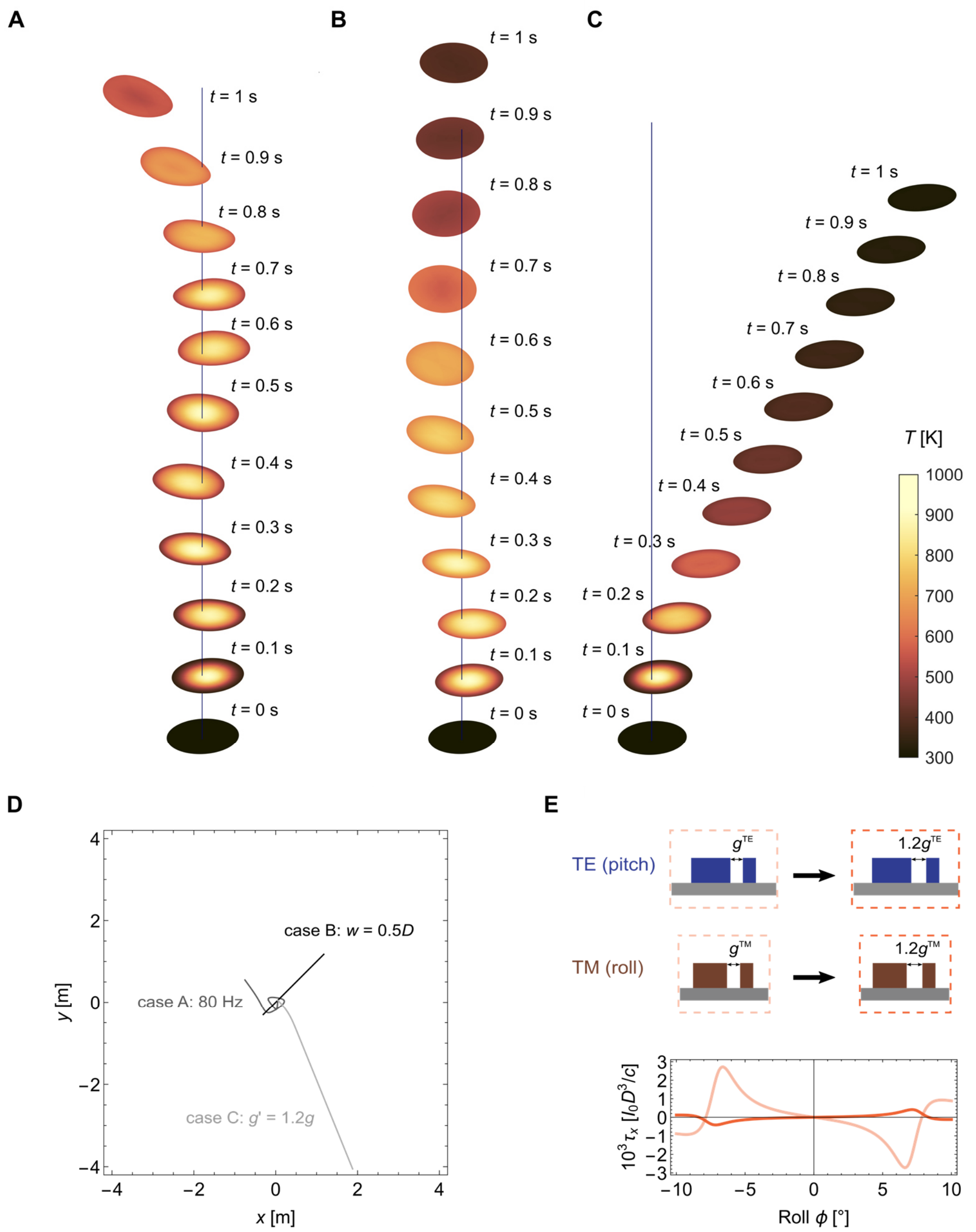

**Fig. S7.** Unstable propulsion of flexible, metagrating-patterned lightsails being initially translated by $x = y = 0.05$ m and tilted by $\theta = \phi = -2°$ due to **(A)** insufficient spin speed $f = 80$ Hz, **(B)** a larger beam width $w = 0.5D$, and **(C)** altered TE and TM metagrating unit cell designs, where the distance between resonators was increased by 20% compared to the stable designs presented in the main text. **(D)** Instability is characterized by the flexible lightsail veering off from the beam center for all three considered cases. **(E)** Specifically, increasing the distance between resonators results in the torque about $x$ changing the sign of its slope and thus losing its roll-restoring ability.

Rounded off to four decimal places, such that the absolute values of the complex eigenvalues, $|\lambda_i|$ for $i = 1, \dots, 8$, are then given by

$$|\lambda_{1,2}| \approx 1.0674, |\lambda_{3,4}| \approx 1, |\lambda_{5,6}| \approx 0.9368, |\lambda_{7,8}| \approx 0.9999.$$

In contrast to the original metagrating design, we see that the absolute values of the eigenvalues do not exclusively lie on the unit circle, indicating an unstable linear time-periodic system, or spinning lightsails that would not be passively (marginally) stabilized.

We note that absolute values of complex eigenvalues for cases A and B ($f$ = 80 Hz and $w = 0.5D$, respectively), are ~1 despite the simulated trajectories being unbounded, indicating that chosen initial conditions lie outside of the linear regime assumed by Floquet theory.

## Supplementary References

# Supplementary Videos

## Dynamically Stable Radiation Pressure Propulsion of Flexible Lightsails for Interstellar Exploration

Ramon Gao[1], Michael D. Kelzenberg[1], and Harry A. Atwater*

California Institute of Technology, Pasadena CA 91125

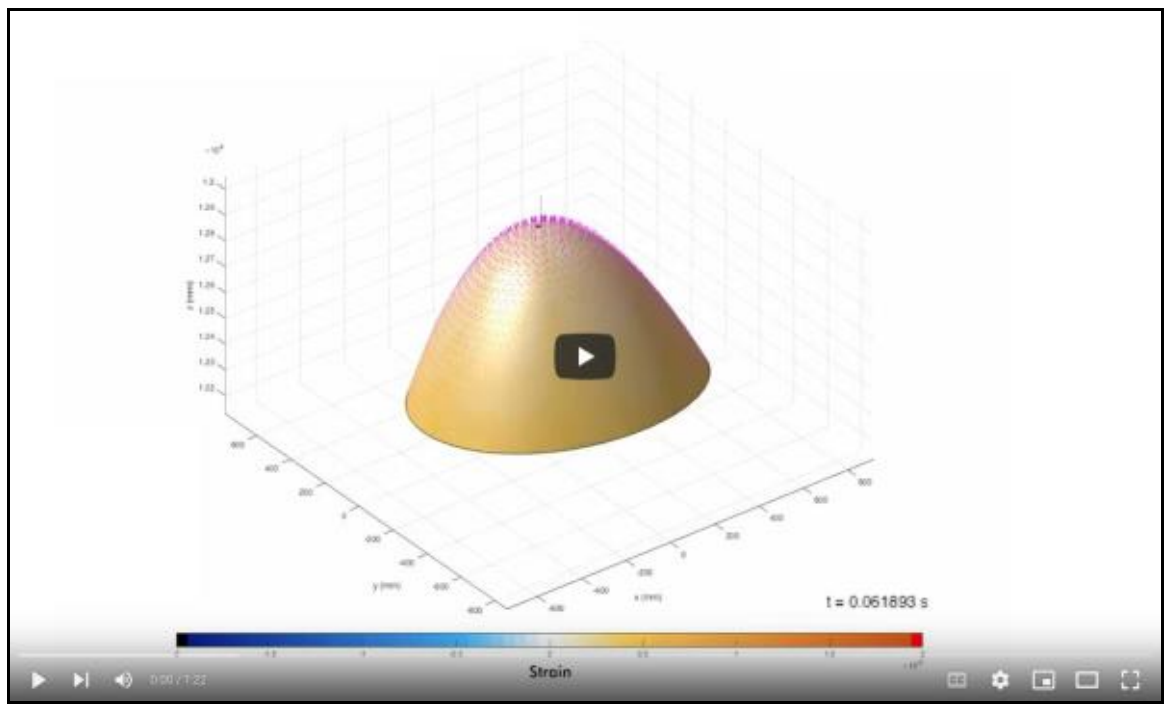

**Video S1.** Simulations of flat and curved specular lightsail dynamics.

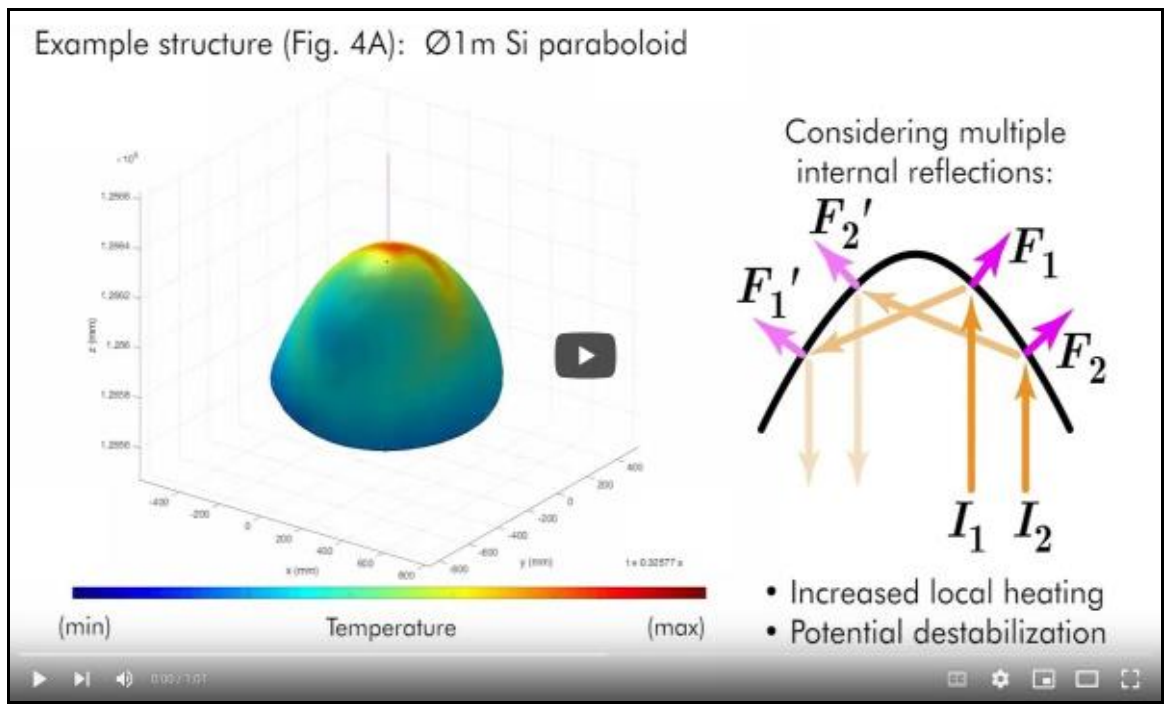

**Video S2.** Raytracing-based simulations of curved specular lightsail dynamics.

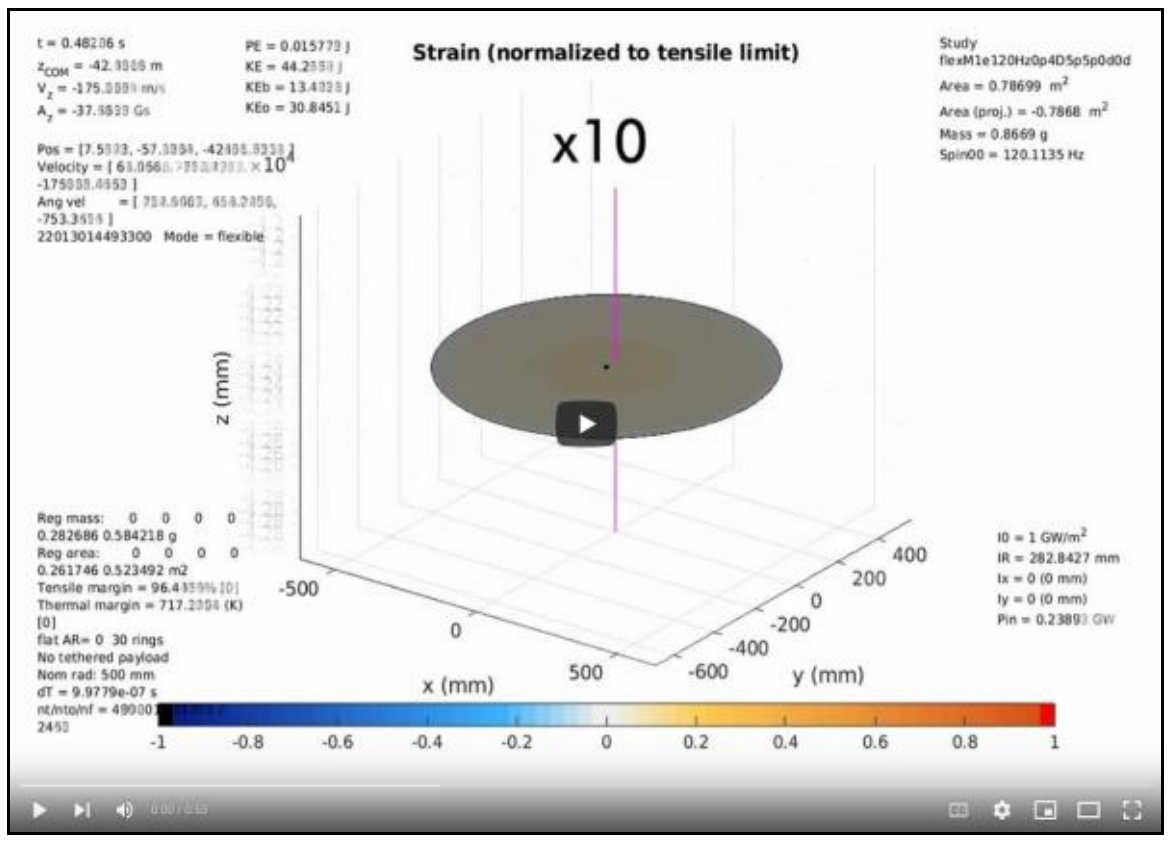

**Video S3.** Simulation of self-stabilizing flat lightsail dynamics based on optical metagratings with initial translational and rotational offset.

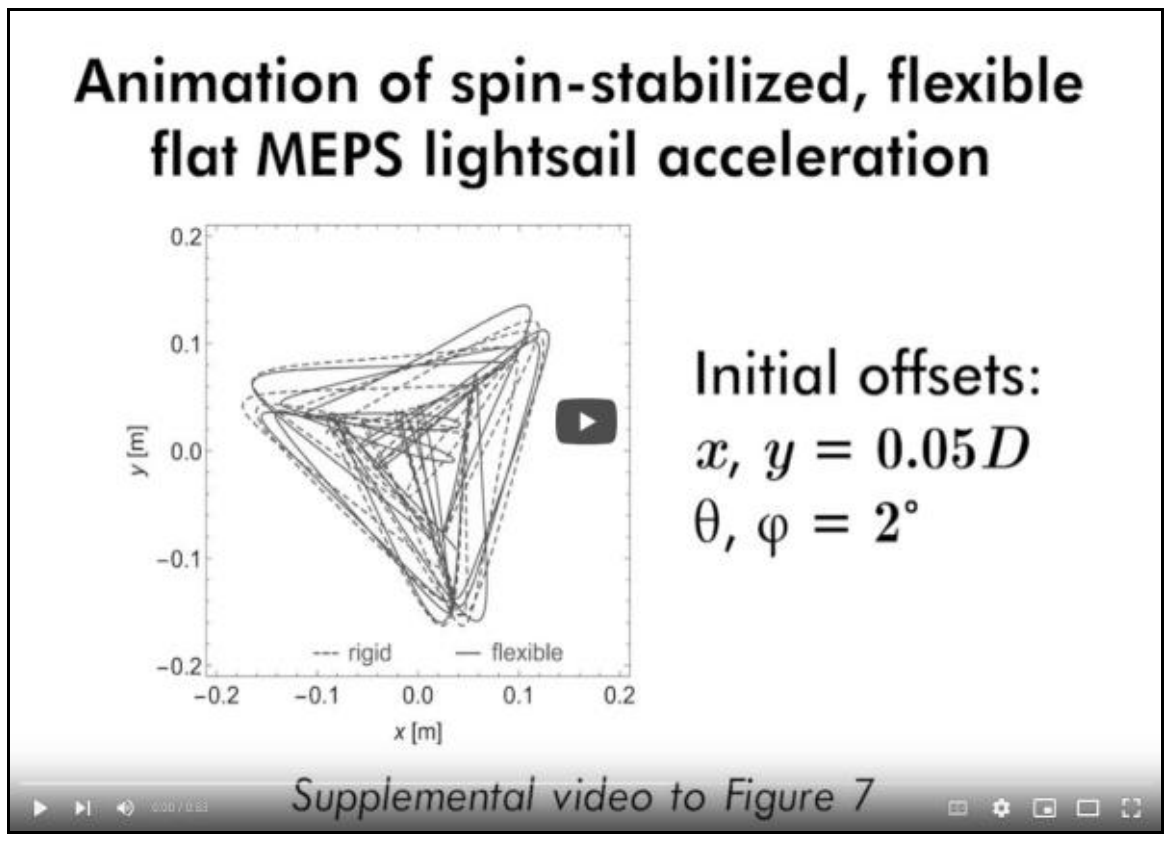

**Video S4.** Simulation of self-stabilizing flat lightsail dynamics based on optical metagratings with initial translational offset only.